\documentclass[10pt]{article}
\usepackage{fullpage}
\usepackage{amsmath}
\usepackage{amssymb}
\usepackage{amsfonts}
\usepackage{comment}
\usepackage{color}
\usepackage{cite}
\usepackage{subcaption}
\usepackage{graphicx}
\usepackage{enumitem}

\def\blue{\textcolor{blue}}

\def\##1{\underline{#1}}
\def\=#1{\underline{\underline{#1}}}

\def\+
#1{\underline{\bf #1}}
\def\*#1{\underline{\underline{\bf #1}}}

\def\r#1{(\ref{#1})}
\def\l#1{\label{#1}}
\def\c#1{\cite{#1}}

\def\le{\left(}
\def\ri{\right)}
\def\les{\left[}
\def\ris{\right]}
\def\lec{\left\{}
\def\ric{\right\}}
\def\lek{[{\kern 0.1em}}
\def\rik{{\kern 0.1em}]}

\def\.{\mbox{ \tiny{$^\bullet$} }}

\def\eps{\varepsilon}

\def\epso{\eps_{\scriptscriptstyle 0}}
\def\lambdao{\lambda_{\scriptscriptstyle 0}}
\def\muo{\mu_{\scriptscriptstyle 0}}

\def\ko{k_{\scriptscriptstyle 0}}

\def\ux{\hat{\#u}_{\rm x}}
\def\uy{\hat{\#u}_{\rm y}}
\def\uz{\hat{\#u}_{\rm z}}

\begin{document}

\begin{center}

\LARGE{ {\bf Characterization of golden vaterite by the extended Maxwell Garnett formalism
}}
\end{center}
\begin{center}
\vspace{10mm} \large

  \vspace{3mm}

 {Tom G. Mackay}\footnote{E--mail: T.Mackay@ed.ac.uk.}\\
{\em School of Mathematics and
   Maxwell Institute for Mathematical Sciences\\
University of Edinburgh, Edinburgh EH9 3FD, UK}\\
and\\
 {\em NanoMM~---~Nanoengineered Metamaterials Group\\ Department of Engineering Science and Mechanics\\
Pennsylvania State University, University Park, PA 16802--6812,
USA}
 \vspace{3mm}\\
 {Akhlesh  Lakhtakia}\\
 {\em NanoMM~---~Nanoengineered Metamaterials Group\\ Department of Engineering Science and Mechanics\\
Pennsylvania State University, University Park, PA 16802--6812, USA}

\normalsize

\end{center}

\begin{center}
\vspace{5mm} {\bf Abstract}
\end{center}

The homogenization of  vaterite
impregnated with gold nanoparticles was accomplished using the extended Maxwell Garnett formalism.
The extended formalism takes into account the intrinsic anisotropy of  vaterite
 as well as the size, shape, and orientation of the nanoparticles. Size-dependent permittivity was used for the gold nanoparticles.
 Numerical studies revealed that the homogenized composite material's  permittivity parameters are acutely sensitive to the size, shape, orientation, and volume fraction of the gold nanoparticles.

 \vspace{5mm}
\vspace{5mm}

\section{Introduction}

Vaterite is a naturally occurring polymorph of calcium carbonate which can be found  in certain biological tissues  \c{Bacterial_vaterite,Vaterite_spherulites,Limpet_shells}.
Vaterite monocrystals can self-assemble to form polycrystalline spherulites which are highly porous \c{Water_Vaterite}. The porosity and biocompatibility of vaterite
spherulites render them attractive platforms for biomedical applications. In particular, the ability to engineer the optical properties of vaterite spherulites 
that are impregnated with
 gold nanoparticles
opens the door to    targeted drug delivery \c{Drug_delivery},  photothermal therapy \c{photothermal},   optical sensing \c{sensing}, etc.

The aim of this study is to rigorously characterize the optical properties of porous vaterite impregnated with gold nanoparticles, using homogenization theory \c{MAEH}. Recently, such a characterization was attempted using the Maxwell Garnett homogenization formalism \c{Noskov}, but the implementation of that formalism  was flawed because: (a) the depolarization factors adopted did not take account of the uniaxial anisotropy of vaterite; (b) the permittivity used for gold nanoparticles did not take account of the size of the nanoparticles; and (c) a simple version of the Maxwell Garnett  formalism was used that did not take into account the size of the inclusions within the vaterite host material. These shortcomings can be avoided, as we show here  by implementing an extended version of the Maxwell Garnett formalism  that
accommodates the size of the   nanoparticles \c{Ch3_M08_JNP} and
  incorporates depolarization factors that   capture the intrinsic uniaxial anisotropy of vaterite \c{Ch3_M97}, using a size-dependent permittivity for the gold nanoparticles \c{Au}.

An $\exp(-i\omega{t})$ dependence on time $t$ is implicit,
with $i=\sqrt{-1}$ and $\omega$ as the angular frequency.
The permittivity and permeability  of free space, respectively, are denoted by
 $\epso$ and $\muo$, whereas
$\lambdao$ and $\ko = 2 \pi / \lambdao$ are   the  wavelength and wavenumber in free space, respectively.
Single underlining denotes a 3-vector, with   $\lec \ux, \uy, \uz \ric$ being the triad of Cartesian vectors. Double underlining with normal typeface denotes a 3$\times$3 dyadic \c{Chen}, with the identity dyadic being $\=I = \ux \, \ux + \uy \, \uy + \uz \, \uz$ and the null dyadic being $\=0$. Double underlining with bold typeface denotes a 6$\times$6 dyadic \c{EAB2}. An $\exp(-i\omega{t})$ dependence on time $t$ is implicit,
with $i=\sqrt{-1}$ and $\omega$ as the angular frequency.
 
 \section{Homogenization}
 
 \subsection{Preliminaries}
 
 A porous host material, labeled \emph{c}, contains spheroidal inclusions. 
Some inclusions  are composed of gold, in which case they are labeled \emph{b}; the remaining inclusions are pores and labeled \emph{a}. 
The volume fraction of    material  $a$ is denoted by $f_a\in[0,1]$ while that of material $b$ by $f_b\in[0,1]$. 

The host material $c$ is vaterite  \c{Noskov}, which is a uniaxial
dielectric material characterized in the visible spectral regime
by an ordinary relative permittivity of 2.4 and an extraordinary relative permittivity of 2.7. The optic axis
of vaterite can vary spatially. However, in any sufficiently large spatial domain in which the optic axis is invariant,
 a Cartesian coordinate system can  be found,
by virtue of the principal axis theorem \c{Strang}, such that
the permittivity dyadic of vaterite can be written as
\begin{equation}
\=\eps_{\,c} = \epso\les2.4 \le \ux \, \ux + \uy\,\uy \ri  + 2.72 \,\uz\,\uz\ris\,
\end{equation}
in that domain. Homogenization
of only that domain is considered here, it being implicit that  the Cartesian coordinate system varies from domain to domain.
This \textit{local} homogenization approach is evident
in  Eq.~(3) of Ref.~\citenum{Noskov}, and it has been
 used in the context of thin films too \cite{LSVE,STFbook}.

\begin{figure}[!htb]
\centering
 \includegraphics[width=6.9cm]{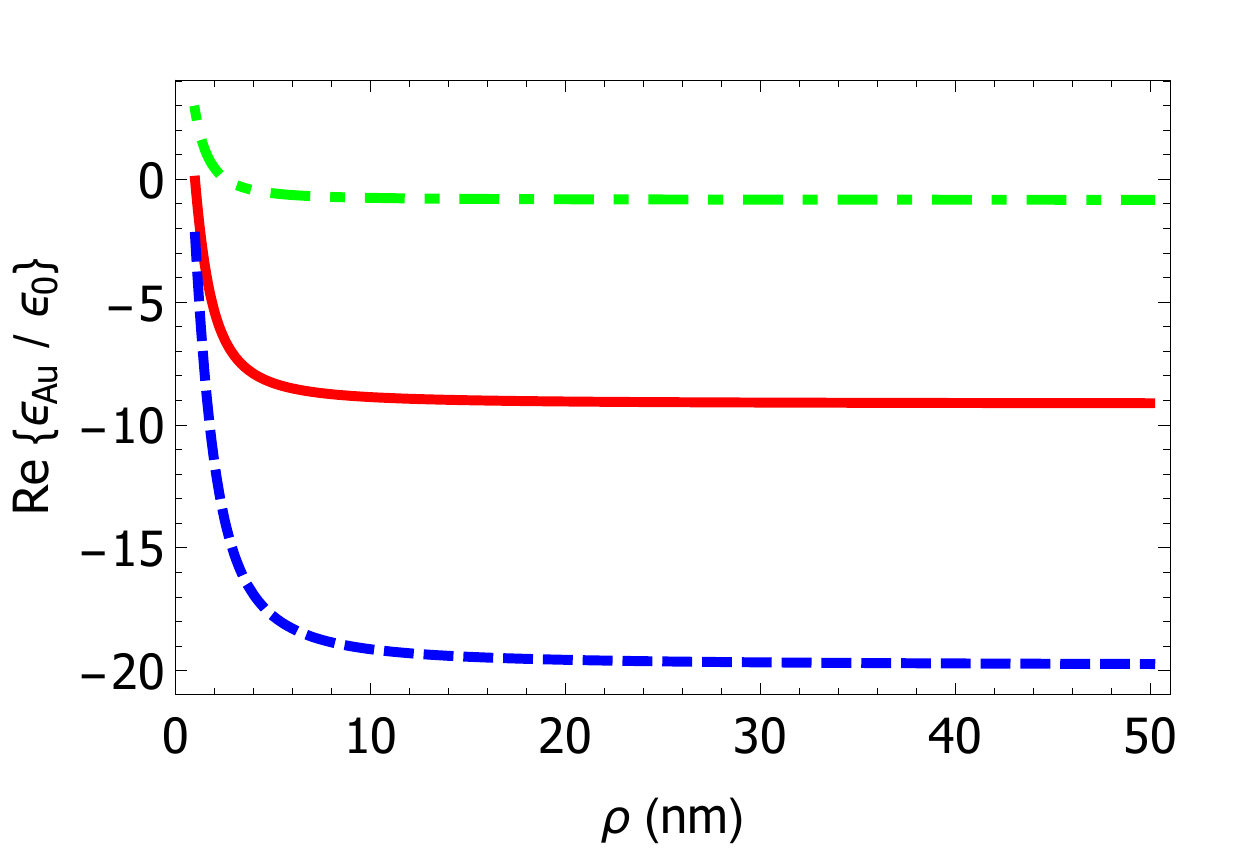} \hfill \includegraphics[width=6.9cm]{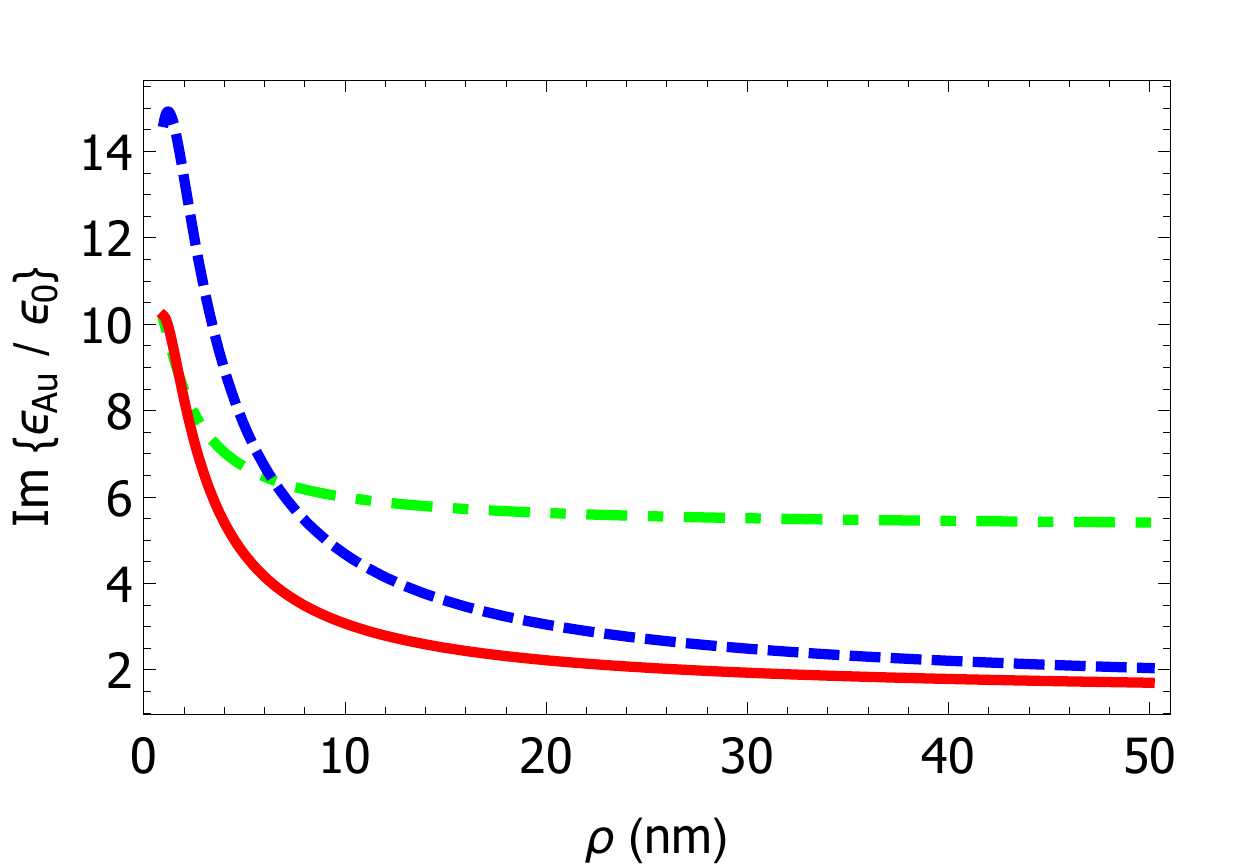} 
 \caption{\label{Fig1} 
 Real and imaginary parts of   the relative permittivity scalar of gold nanoparticles of radius $\rho$ for free-space wavelengths $\lambdao = 450$ nm  (green, broken dashed curve), 600 nm (red, solid curve), and 750 nm (blue, dashed curve).
}
\end{figure}

The permittivity of inclusion material $a$ is simply that of free space, i.e., $\eps_a = \epso$.
For  material $b$, we consider the
  the  permittivity of  gold nanoparticles of radius $\rho$ at the angular frequency $\omega$, which
is provided by the formula \c{Au}
\begin{equation} \l{Au}
\eps_{Au} = \epso \le \eps_{Di} + i \Delta \eps \ri,
\end{equation}
 where
 \begin{equation} \l{Au2}
 \left.
 \begin{array}{l}
 \eps_{Di} = \displaystyle{\eps_r - \frac{\omega^2_p}{\omega^2 + i \le \gamma_{bulk} + C \frac{v_F}{\rho} \ri \omega}} \vspace{6pt}\\
 \Delta \eps = \displaystyle{ \frac{A}{1 + \exp \les - \frac{\omega - \omega_c}{\Delta}\ris}}
\end{array}
\right\},
 \end{equation}
 with $\eps_r = 9.84$,  $A= 5.6$, $C= 0.33$,  $\omega_c = 3.6462 \times 10^{15}$~rad~s$^{-1}$, $\Delta = 2.5828 \times 10^{14}$~rad~s$^{-1}$,  $\omega_p = 1.3689 \times 10^{16}$~rad~s$^{-1}$, $\gamma_{bulk} = 1.0939 \times 10^{14}$~rad~s$^{-1}$, and
 $v_F = 1.4 \times 10^6 \, \mbox{m s}^{-1}$.  
 The real and imaginary parts of $\eps_{Au}/\epso$ are plotted against $\rho$ in Fig.~\ref{Fig1} for $\lambdao \in\lec 450, 600,750\ric$~nm. 
 Clearly, $\eps_{Au}$ is acutely sensitive to both the nanoparticle size, especially over the range $ \rho < 25 $ nm,  and the free-space wavelength. Specifically, $\mbox{Re}\lec \eps_{Au} \ric$ becomes more negative as $\rho$ increases and as $\lambdao$ increases, while $\mbox{Im}\lec \eps_{Au} \ric$ becomes more positive as $\rho$ decreases.
With the assumption that Eq.~\r{Au} extends to 
   gold nanoparticles  of spheroidal shape that is neither very prolate nor very oblate, we set $\eps_b = \eps_{Au}$. 

To begin with, let all inclusions have same shape,    size, and orientation denoted via the dyadic
 \begin{equation} \l{U_dyadic}
 \=U = \rho \, \=S(\gamma, \beta, \psi) \. \=U_{\,0} \. \les\=S(\gamma, \beta, \psi) \ris^{-1}.
 \end{equation}
 Later (in Fig.~\ref{Fig7}) the requirement that all inclusions have the same orientation is relaxed. 
 The inclusion shape is prescribed by the diagonal dyadic
 \begin{equation} \l{U0}
 \=U_{\,0} = \frac{1}{\sqrt[3]{U}}\le U \ux \, \ux + \uy\,\uy + \uz \, \uz \ri,
 \end{equation}
with shape parameter $U>0$; the size parameter $\rho > 0$ 
provides a measure of the  linear dimensions of the inclusion; and the  orientation is prescribed by the orthogonal dyadic
\begin{equation}
\=S (\gamma, \beta, \psi) = \=R_{\,\text{z}} (\gamma) \. \=R_{\,\text{y}} (\beta) \. \=R_{\,\text{z}} (\psi),
\end{equation}
wherein the rotation dyadics
\begin{equation}
\left.
\begin{array}{l}
\=R_{\,\text{z}} (\kappa) = \le \ux \, \ux + \uy\,\uy \ri\cos \kappa  +  \le \uy \, \ux - \ux\,\uy \ri \sin \kappa \vspace{8pt} \\ \hspace{70mm}+\uz\,\uz\,
\vspace{8pt}
\\
 \=R_{\,\text{y}} (\beta)  = \le \ux \, \ux + \uz\,\uz \ri\cos \beta  +  \le \uz \, \ux - \ux\,\uz \ri \sin \beta \vspace{8pt} \\ \hspace{70mm}+\uy\,\uy\,
\end{array}
\right\},
\end{equation}
  with $\gamma$, $\beta$, and $\psi$ being the Euler angles. Thus, the inclusions are arbitrarily oriented relative to the optic axis of the vaterite host material.

 \subsection{Extended Maxwell Garnett formalism}

Provided that $\lambdao$ is much larger than $\rho$,  porous vaterite  embedded with gold nanoparticles   may be regarded as a homogeneous composite material (HCM).   The permittivity dyadic of the HCM, denoted by $\=\eps_{\,HCM}$, is estimated using an extended version of the  Maxwell Garnett formalism that accommodates the sizes of the inclusions.

The extended Maxwell Garnett formalism \cite{L-emg,Prinkey}
accommodates the sizes of the inclusions whereas
 the standard Maxwell Garnett formalism does not. This is achieved by
  taking both singular and nonsingular contributions  into account in the integration of the corresponding 
dyadic Green function \c{Fikioris,Wang}.
   The extended Maxwell Garnett estimate of the permittivity dyadic of the HCM is $\=\eps_{\,MG}$. Thus, we  have $\=\eps_{\,HCM} \approx  \=\eps_{\,MG}$ with
    \c{MAEH}
   \begin{eqnarray} \l{epsMG}
   \=\eps_{\,MG} &=& \=\eps_{\,c} + f_a\, \=\alpha_{\,a/c} \. \le \=I - i \omega f_a \=D_{\,I,c} \. \=\alpha_{\,a/c} \ri^{-1} \nonumber \\ && +
   f_b \, \=\alpha_{\,b/c} \. \le \=I - i \omega f_b \=D_{\,I,c} \. \=\alpha_{\,b/c} \ri^{-1},
   \end{eqnarray}
   wherein the polarizability density dyadics
   \begin{eqnarray}
   \=\alpha_{\,\ell/c} &=& 
   \le \eps_{\,\ell}\, \=I - \=\eps_{\,c} \ri \. \les \=I + i \omega \=D_{\,U,c}
   \. \le \eps_{\,\ell}\, \=I - \=\eps_{\,c} \ri \ris^{-1}, \nonumber \\ && \ell \in\lec a, b\ric\,.
   \end{eqnarray}
 The size-dependent depolarization dyadic $\=D_{\,U,c}$ is associated  with a cavity of shape specified by $\=U$  in the host material $c$. The
 size-dependent depolarization dyadic $\=D_{\,I,c}$ is equivalent to $\=D_{\,U,c}$ evaluated for $\=U = \=I$. 
 An integral expression for $\=D_{\,U,c}$ is provided in the Appendix. 
 
Since the rotational symmetry axis of the   spheroidal inclusions $a$ and $b$ is not aligned with the optic axis of the host material $c$, the HCM has biaxial symmetry.
In order to focus on inclusion orientation  in the $xy$ plane, as prescribed by the angle $\psi$, we fix $\gamma = \beta = 0$.
 Accordingly,  the HCM's permittivity dyadic has the general form
\begin{eqnarray}
  \=\eps_{\,MG}  &=& \eps^{MG}_x \ux\,\ux + \eps^{MG}_y \uy\,\uy + \eps^{MG}_z \uz\,\uz \nonumber \\ &&+ \eps^{MG}_t \le \ux\,\uy + \uy \, \ux \ri,
\end{eqnarray}
with the off-diagonal permittivity  parameter $ \eps^{MG}_t $ being null valued when the  orientation angle $\psi = n \pi /2$,
$n\in\lec0, 1, 2, 3\ric$.

 \subsection{Numerical results}
We now numerically illustrate the dependency of  the HCM's permittivity parameters $\eps^{MG}_{x,y,z,t}$
 on the shape, size, orientation, and volume fraction of the inclusions, for $\lambdao \in\lec 450, 600,750\ric$~nm.
We take  $f_a + f_b = 0.3$, which is representative of vaterite spherulites. 

\begin{figure}[!htb]
\centering
 \includegraphics[width=6.9cm]{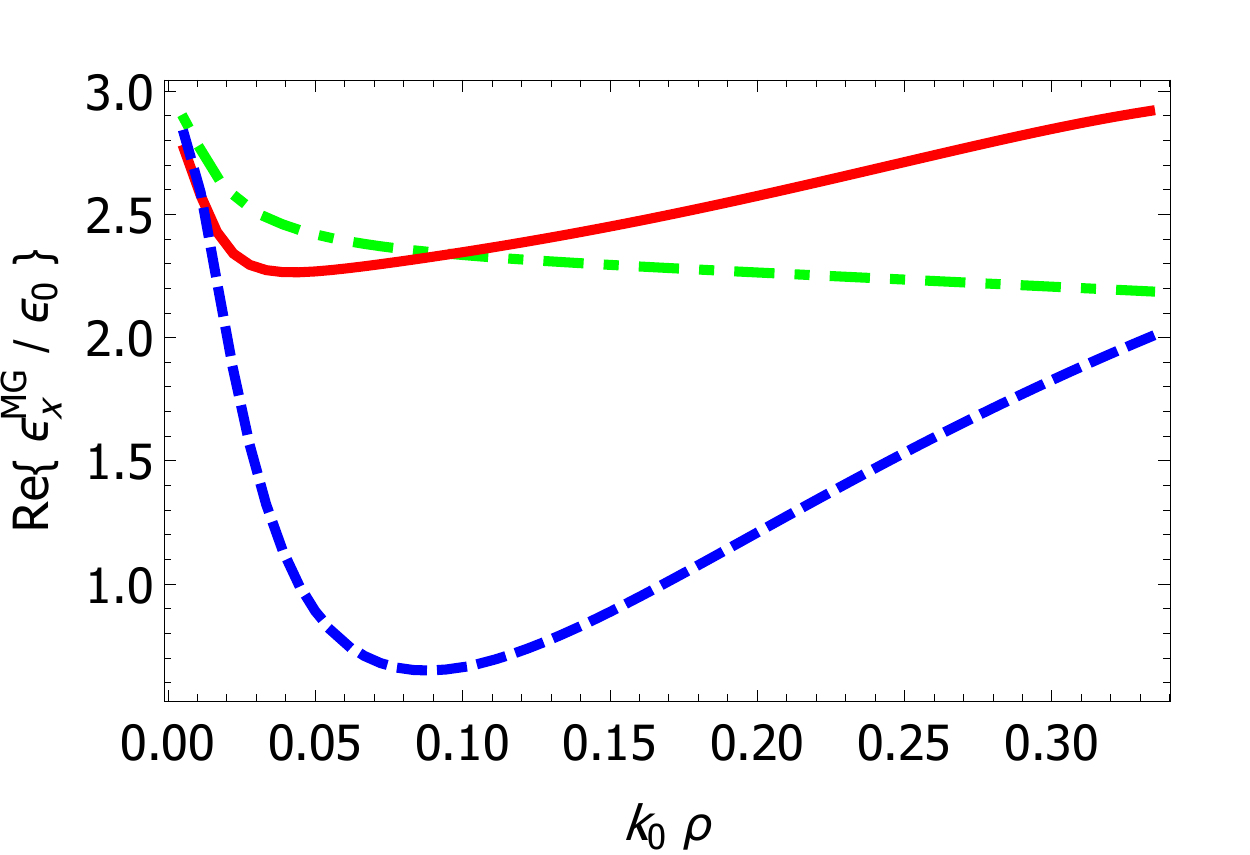} \hfill  \includegraphics[width=6.9cm]{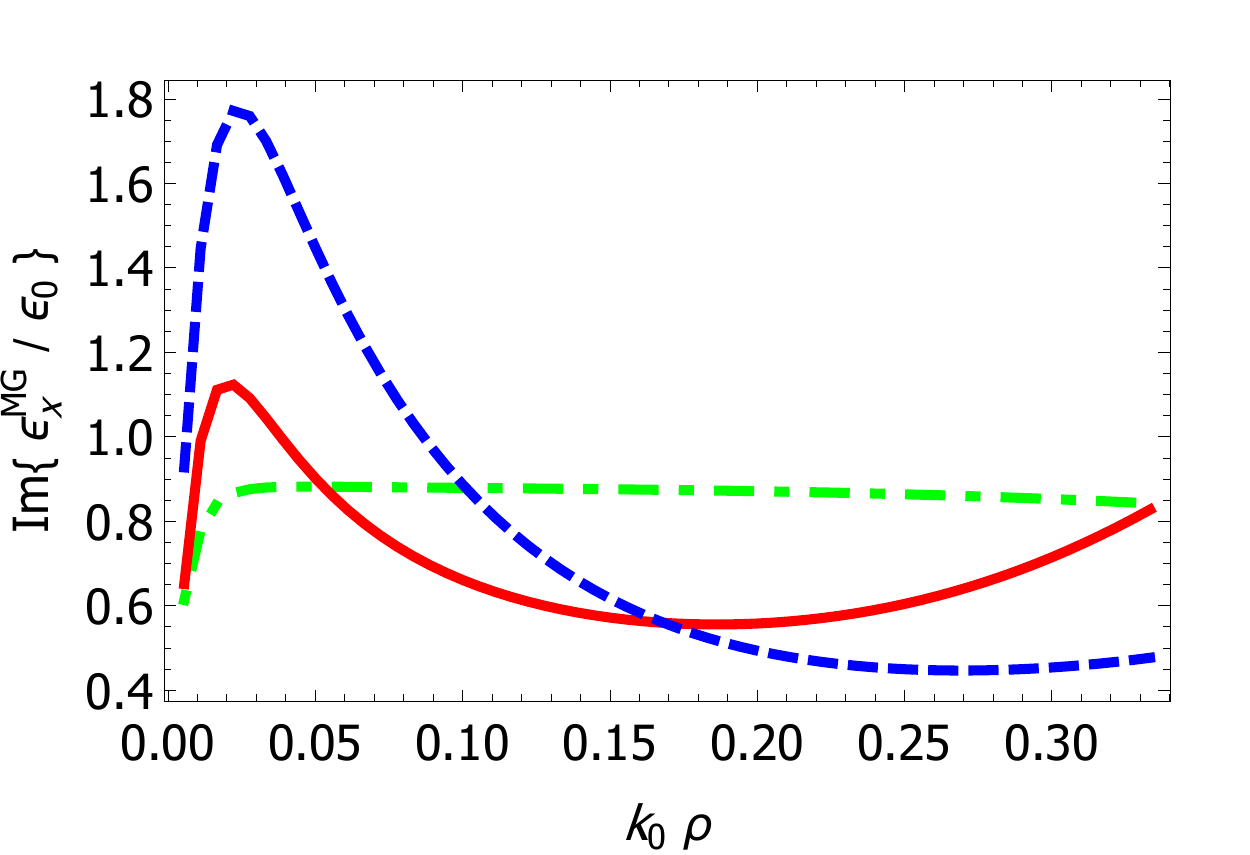} \\
 \includegraphics[width=6.9cm]{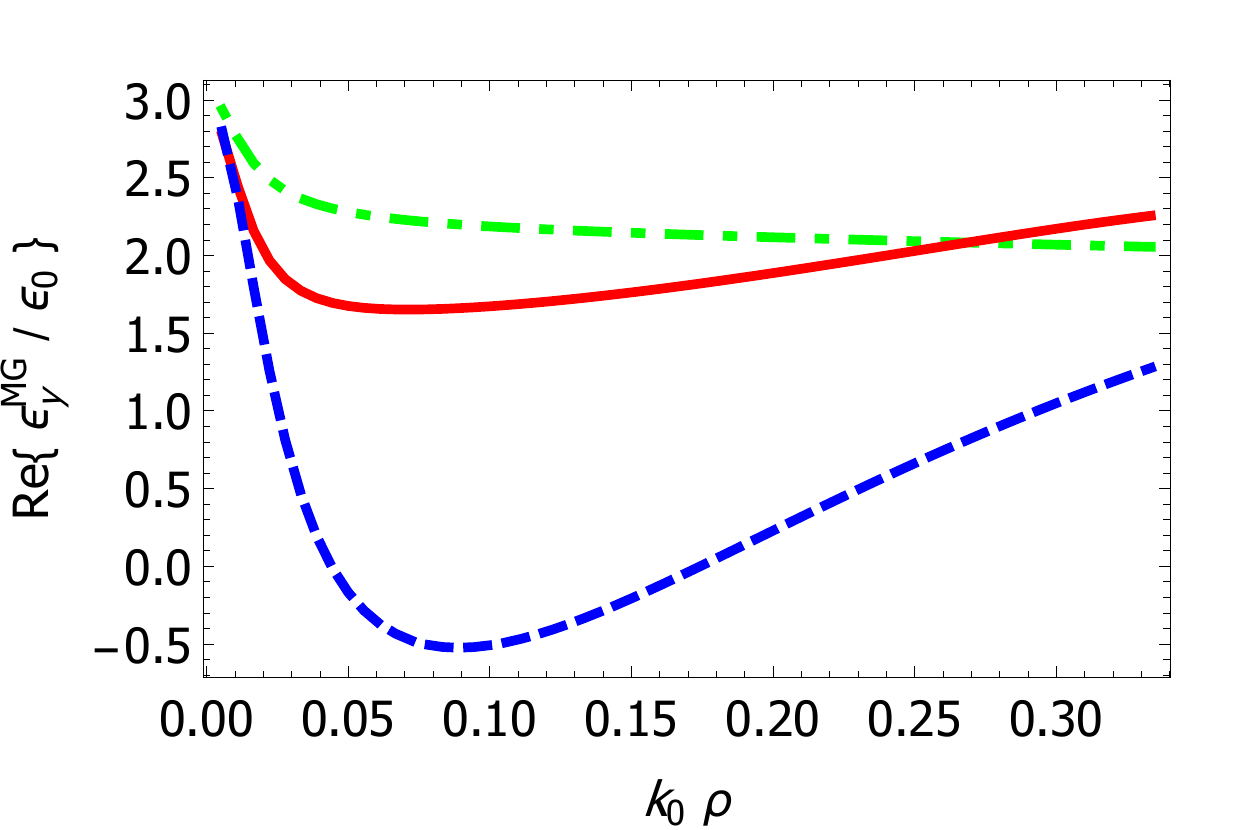} \hfill  \includegraphics[width=6.9cm]{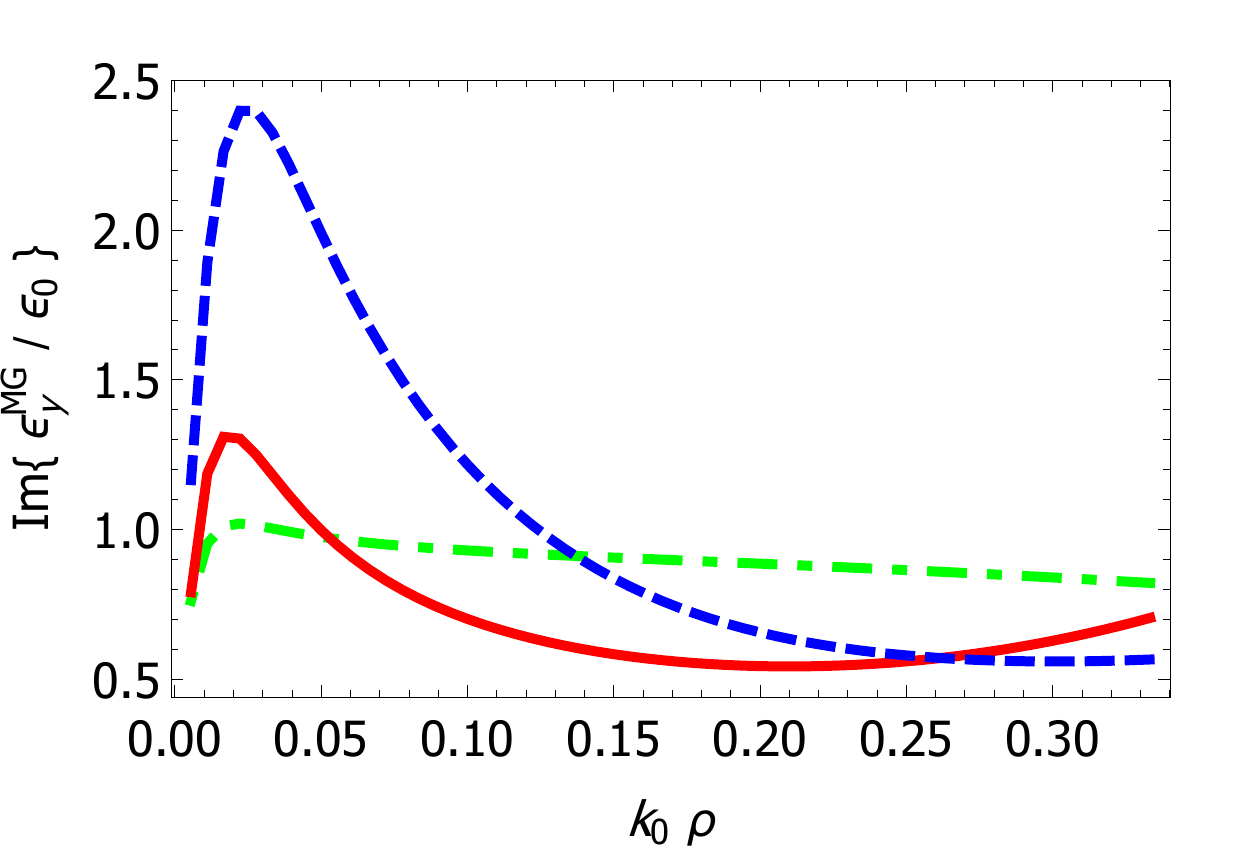} \\
 \includegraphics[width=6.9cm]{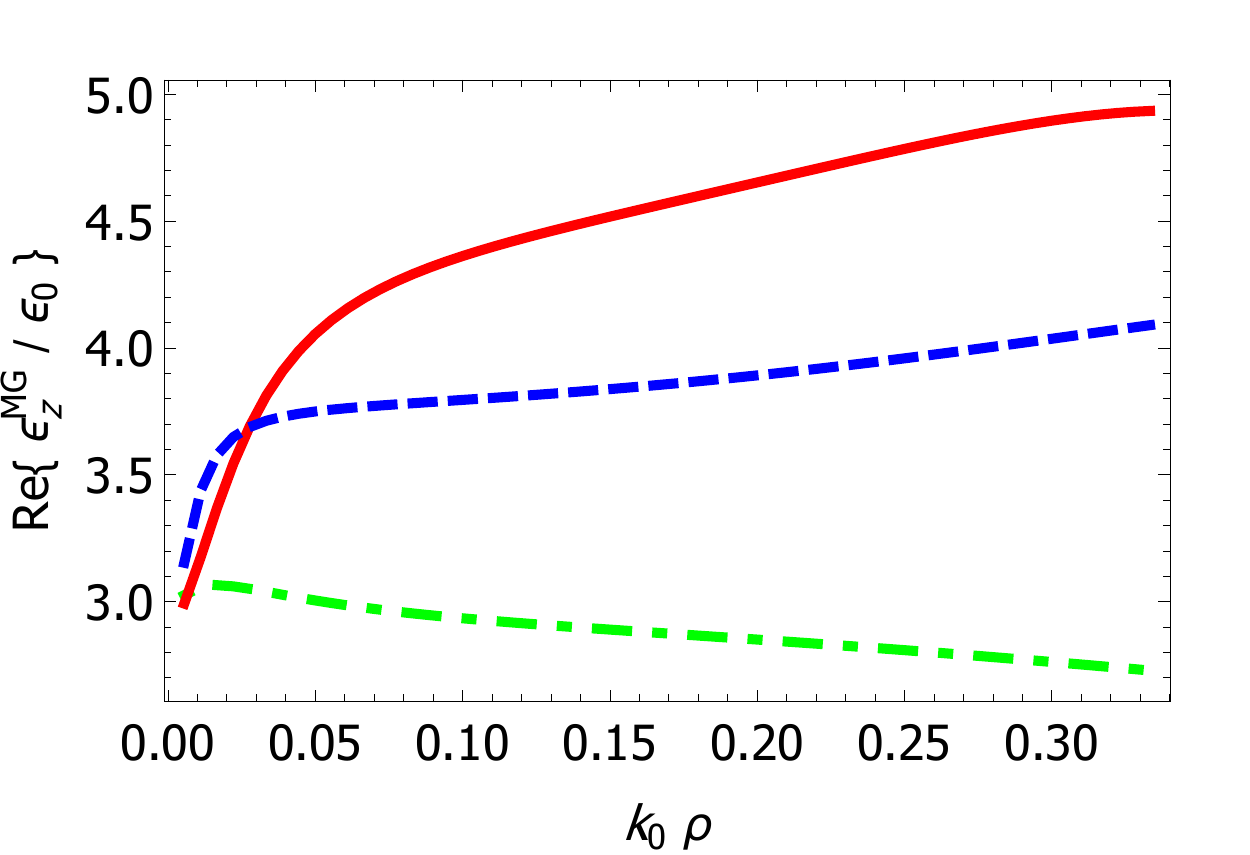} \hfill  \includegraphics[width=6.9cm]{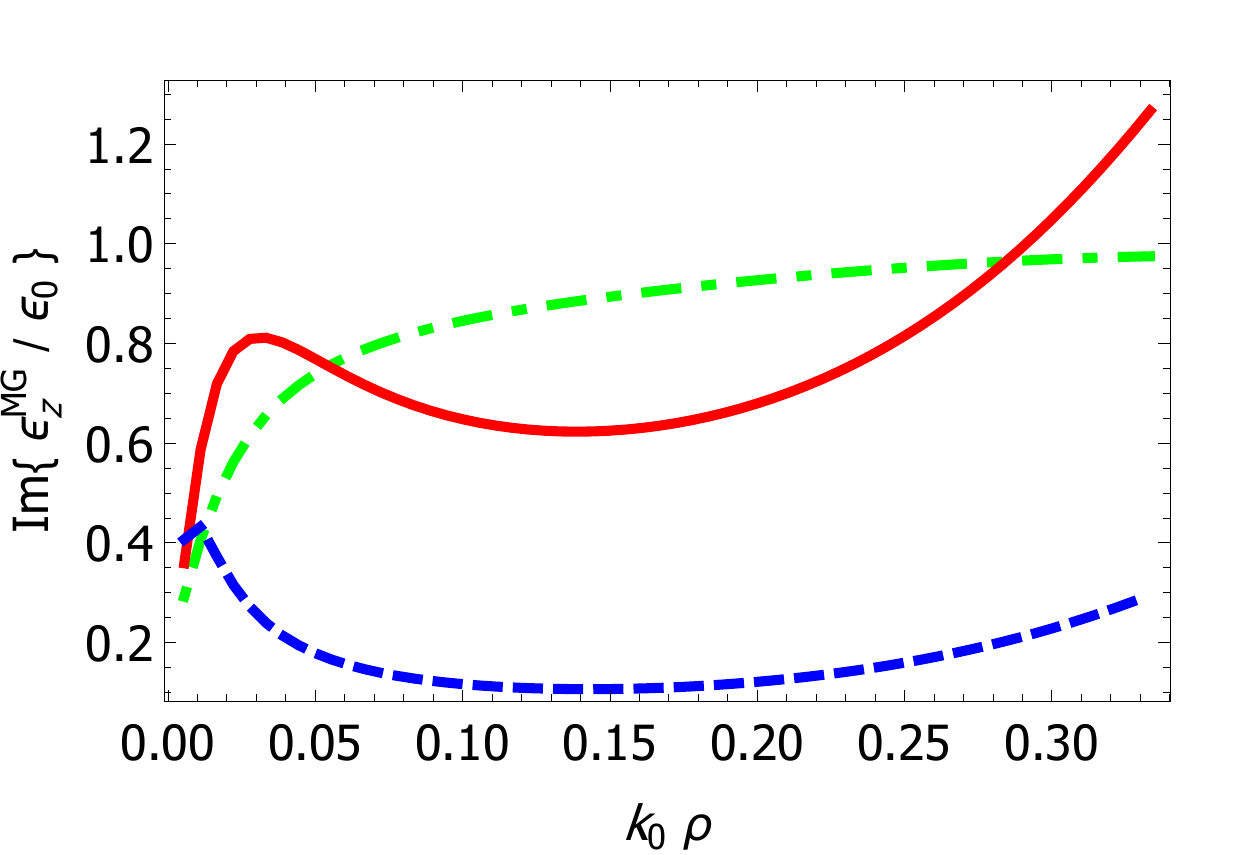} \\
 \includegraphics[width=6.9cm]{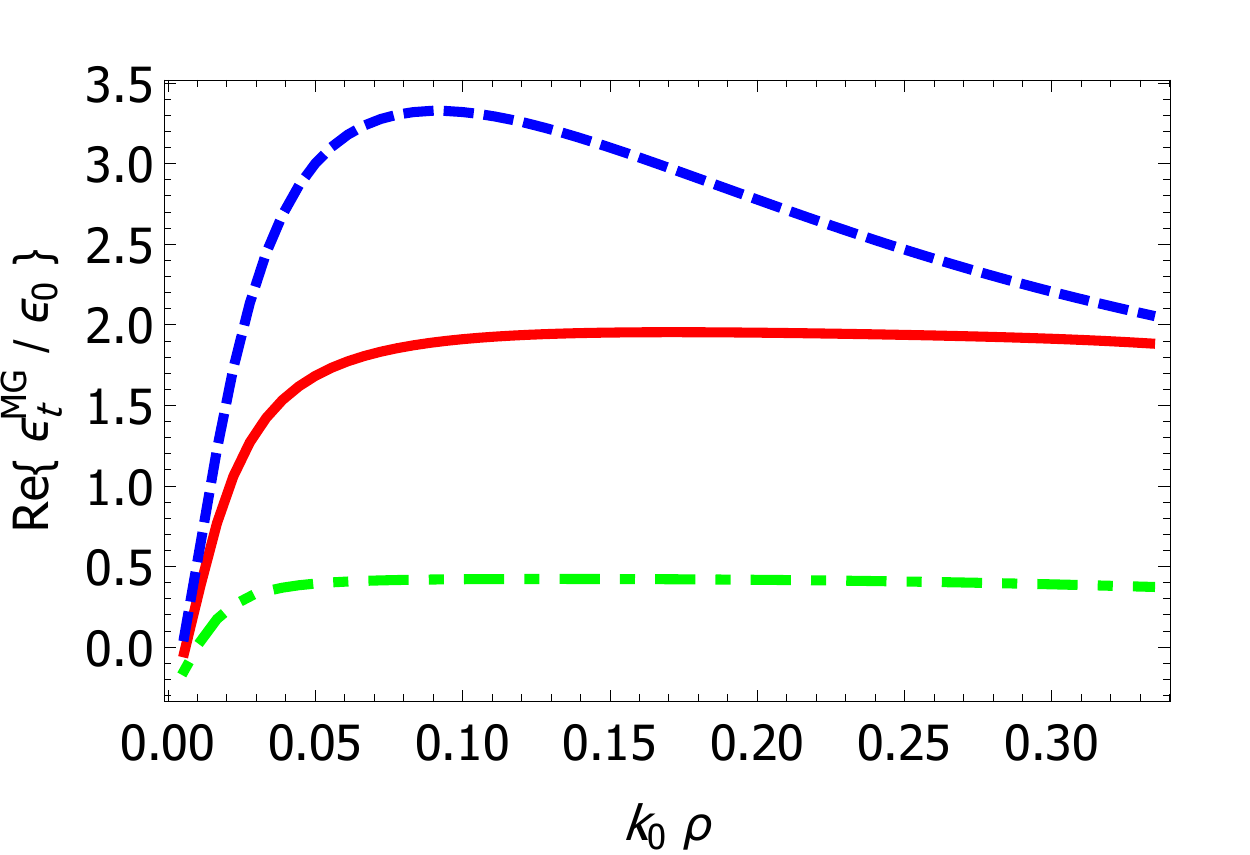} \hfill  \includegraphics[width=6.9cm]{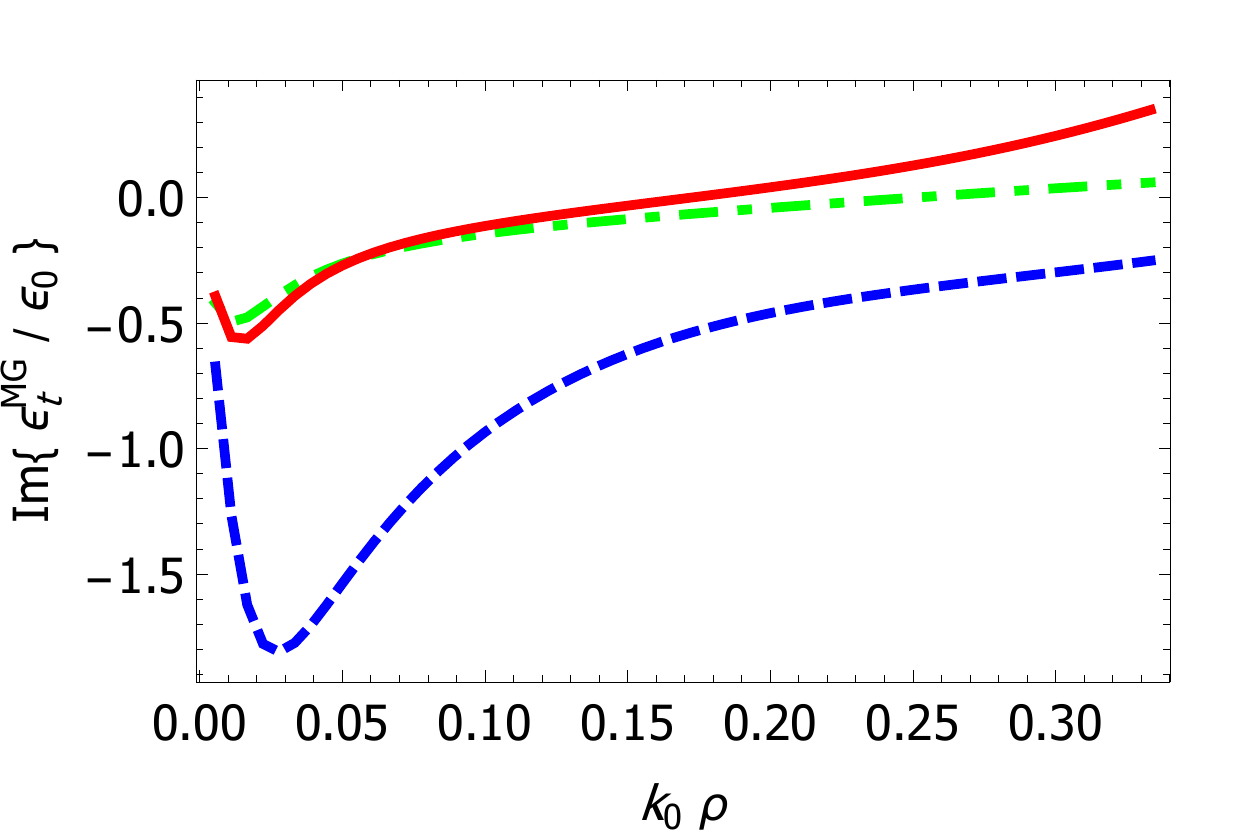}
 \caption{\label{Fig2} 
 Real and imaginary parts of  the non-zero components of the HCM's relative permittivity dyadic   
 versus relative size of gold nanoparticles for free-space wavelengths $\lambdao = 450$ nm  (green, broken dashed curve), 600 nm (red, solid curve), and 750 nm (blue, dashed curve). Shape parameter $U=3$, orientation angle 
 $\psi = 50^\circ$, and volume fractions $f_a = f_b = 0.15$.
}
\end{figure}

Let us begin with dependency on the size parameter $\rho$. In Fig.~\ref{Fig2}, the real and imaginary parts of $\eps^{MG}_{x,y,z}/\epso$ are plotted against $\ko \rho$, as computed using Eq.~\r{epsMG} with $U=3$, $\psi = 50^\circ $, and $f_a = f_b = 0.15$. Both real and imaginary parts
 of $\eps^{MG}_{x,y,z,t}/\epso$
 are acutely sensitive to $\rho$, for all free-space wavelengths considered. 
 The plots for both the real and imaginary parts of $\eps^{MG}_{x}/\epso$ and $\eps^{MG}_{y}/\epso$
 are qualitatively similar for all wavelengths considered.
  For $\lambdao \in\lec  600,750\ric$~nm,   
 the plots of  $\mbox{Im}\lec \eps^{MG}_{x,y,z}/\epso \ric$   exhibit a distinctive local maximum at small values of $\rho$ whereas the plots  of  $\mbox{Im}\lec \eps^{MG}_{t}/\epso \ric$   exhibit a distinctive local minimum at small values of $\rho$.
 For $\lambdao \in\lec  600,750\ric$~nm,  
 the plots of $\mbox{Re}\lec \eps^{MG}_{x,y}/\epso  \ric$ exhibit a distinctive local minimum at small values of $\rho$ whereas
 the plots of $\mbox{Re}\lec \eps^{MG}_{t}/\epso  \ric$ exhibit a distinctive local maximum at small values of $\rho$.
The real and imaginary parts of $\eps^{MG}_{x,y,z,t}/\epso$ are highly sensitive to $\ko \rho$ for the entire range of nanoparticle size considered, but in Fig.~\ref{Fig1} we see that the  permittivity of gold nanoparticles
is  highly sensitive to $\rho$ only for  $\rho \lessapprox \blue{25}$ nm. Therefore, the sensitivity exhibited in Fig.~\ref{Fig2} for large values of $\ko \rho$ is attributable 
to the size dependences of the depolarization dyadics $\=D_{\,U,c}$ and  $\=D_{\,I,c}$ (as appropriate).

\begin{figure}[!htb]
\centering

 \includegraphics[width=6.9cm]{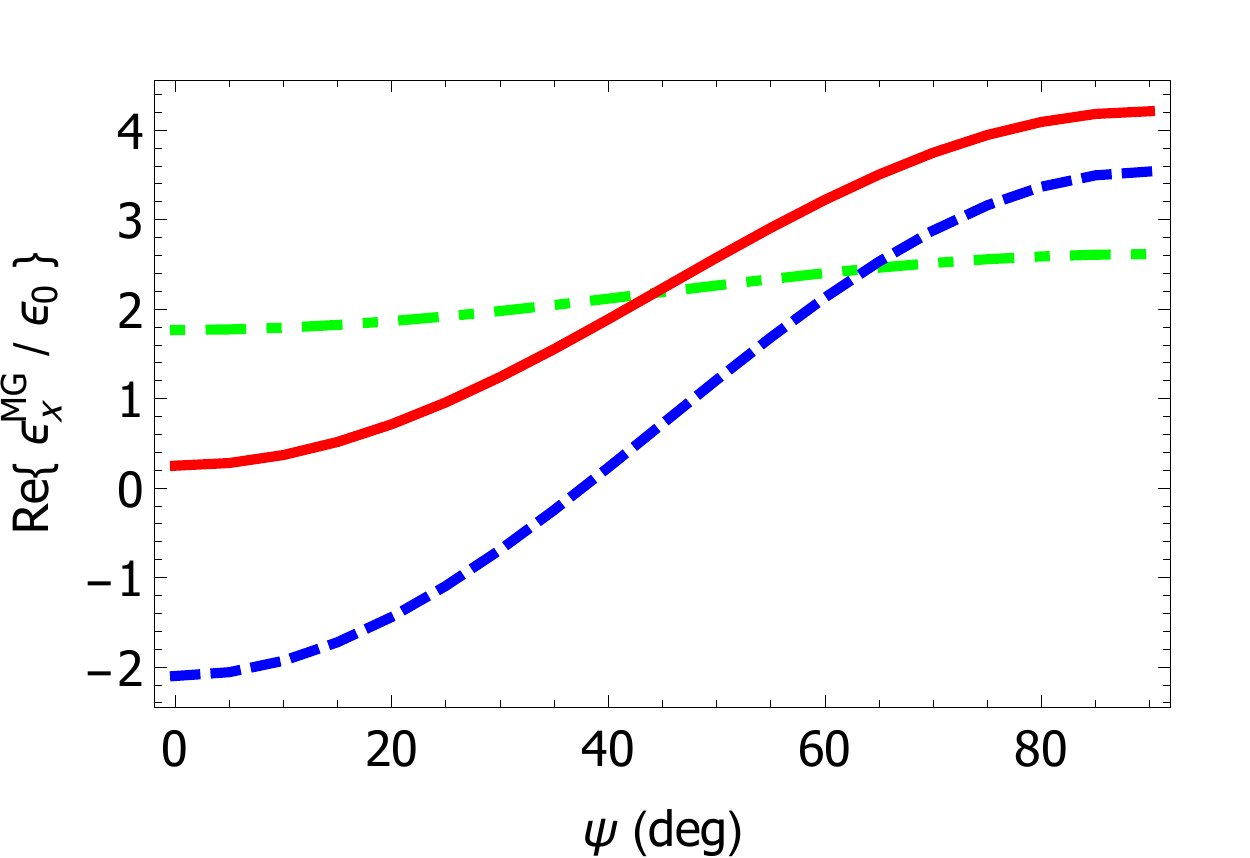} \hfill  \includegraphics[width=6.9cm]{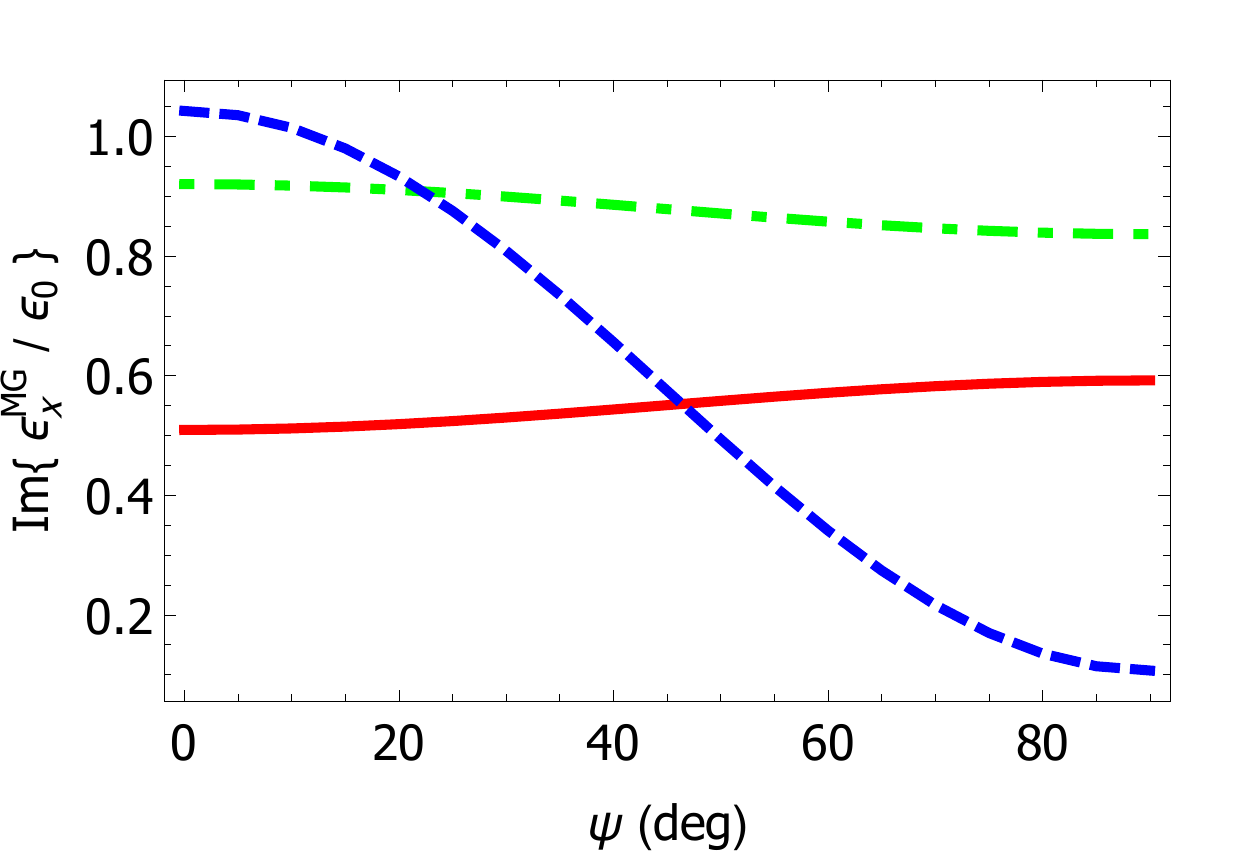} \\
 \includegraphics[width=6.9cm]{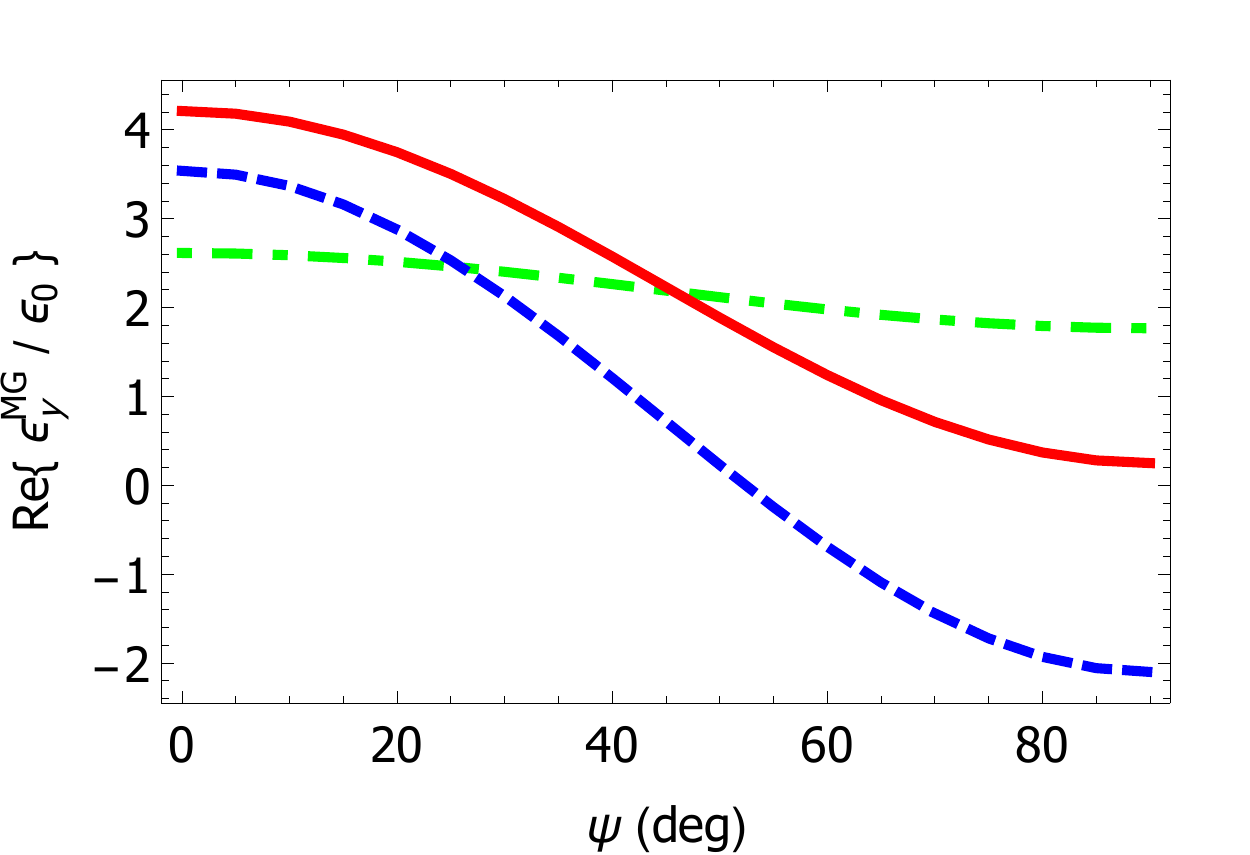} \hfill  \includegraphics[width=6.9cm]{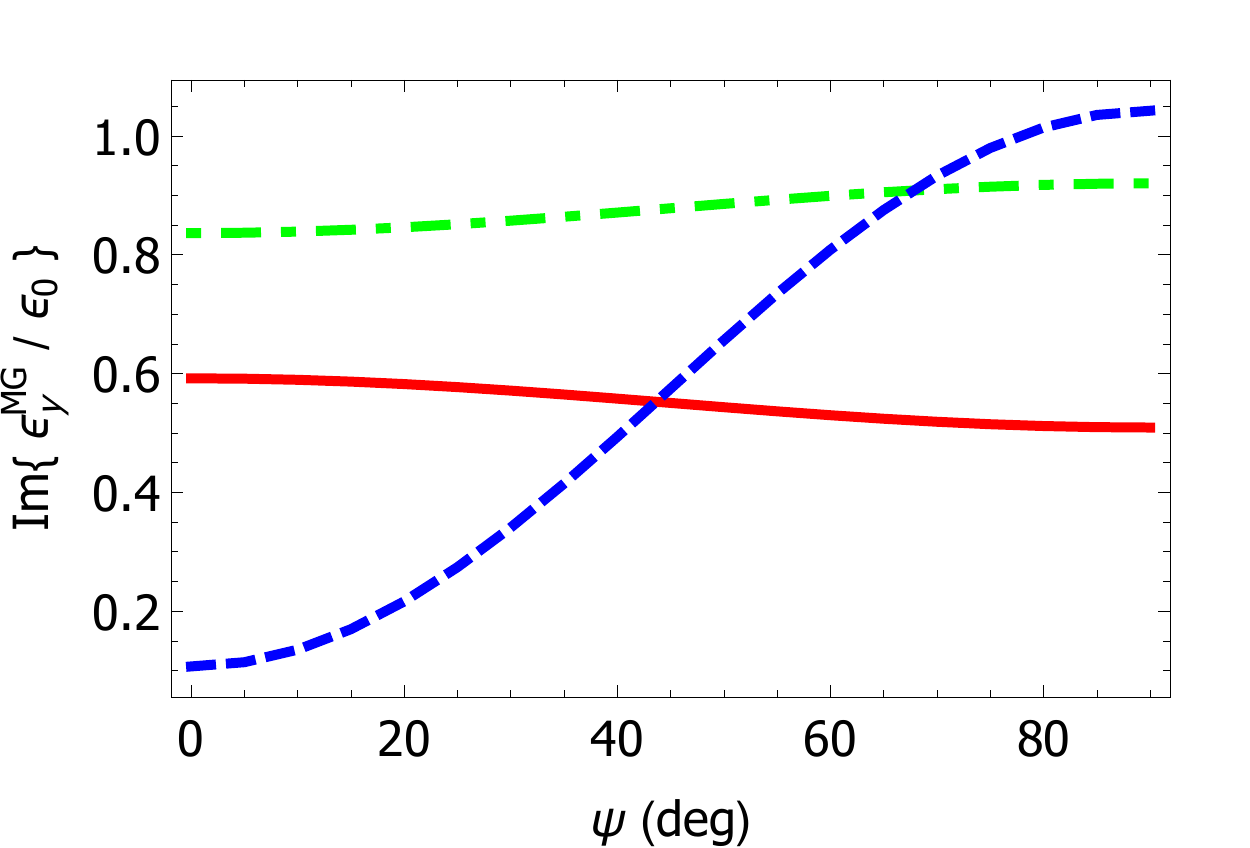} \\
 \includegraphics[width=6.9cm]{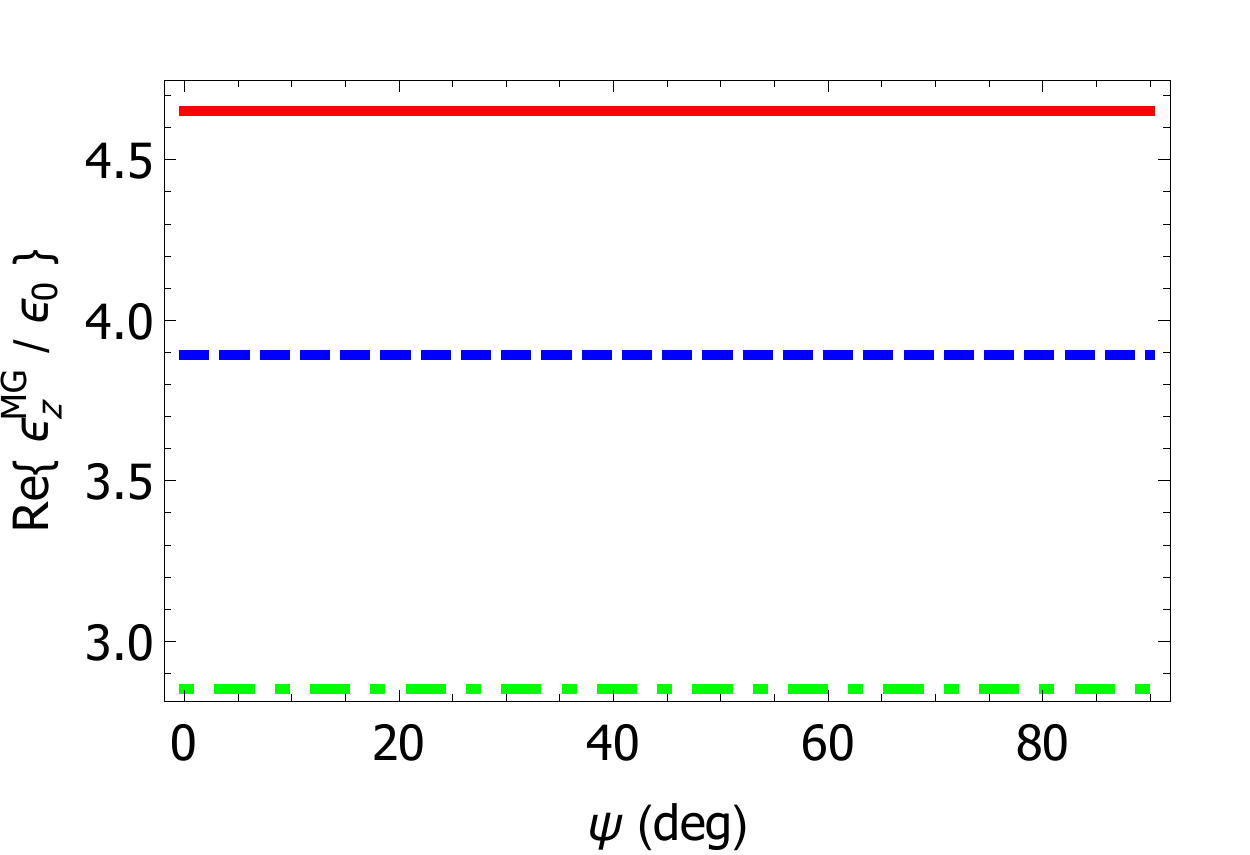} \hfill  \includegraphics[width=6.9cm]{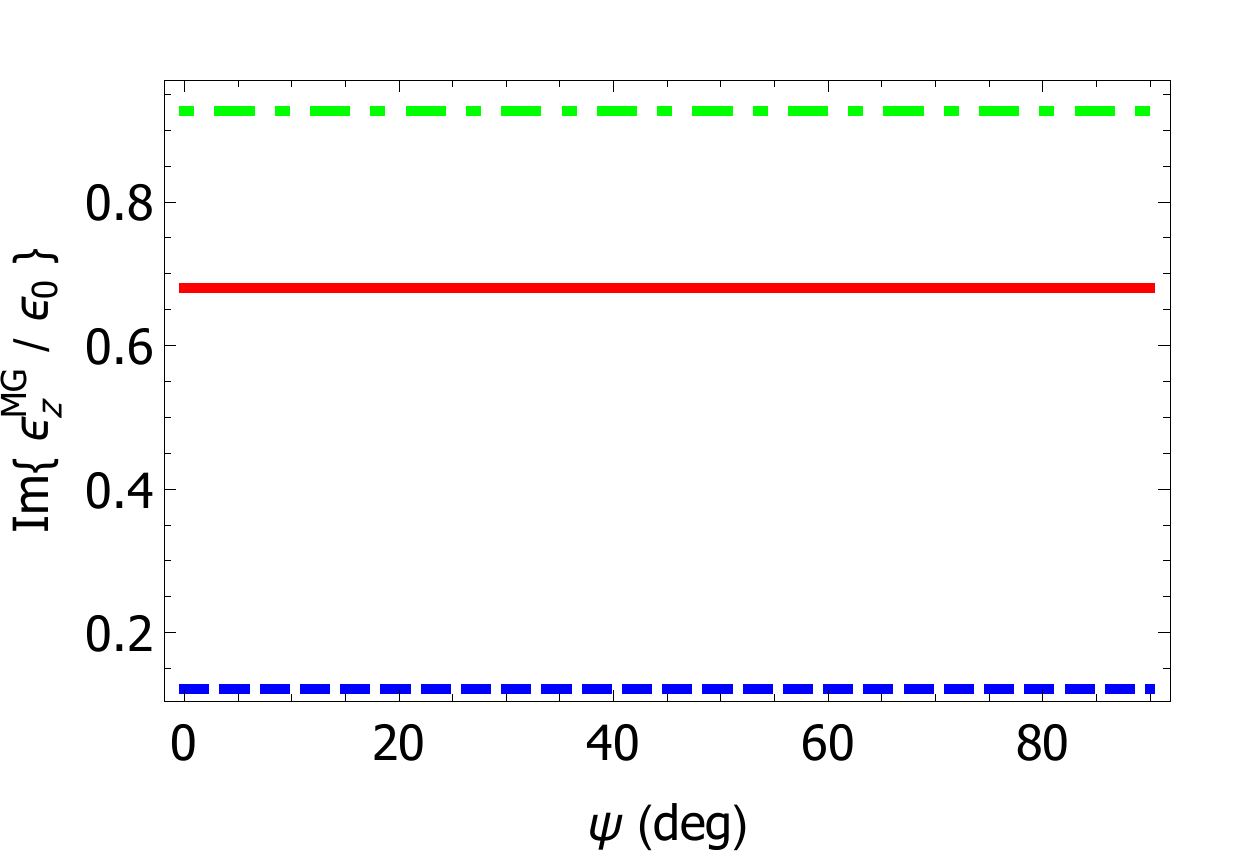} \\
 \includegraphics[width=6.9cm]{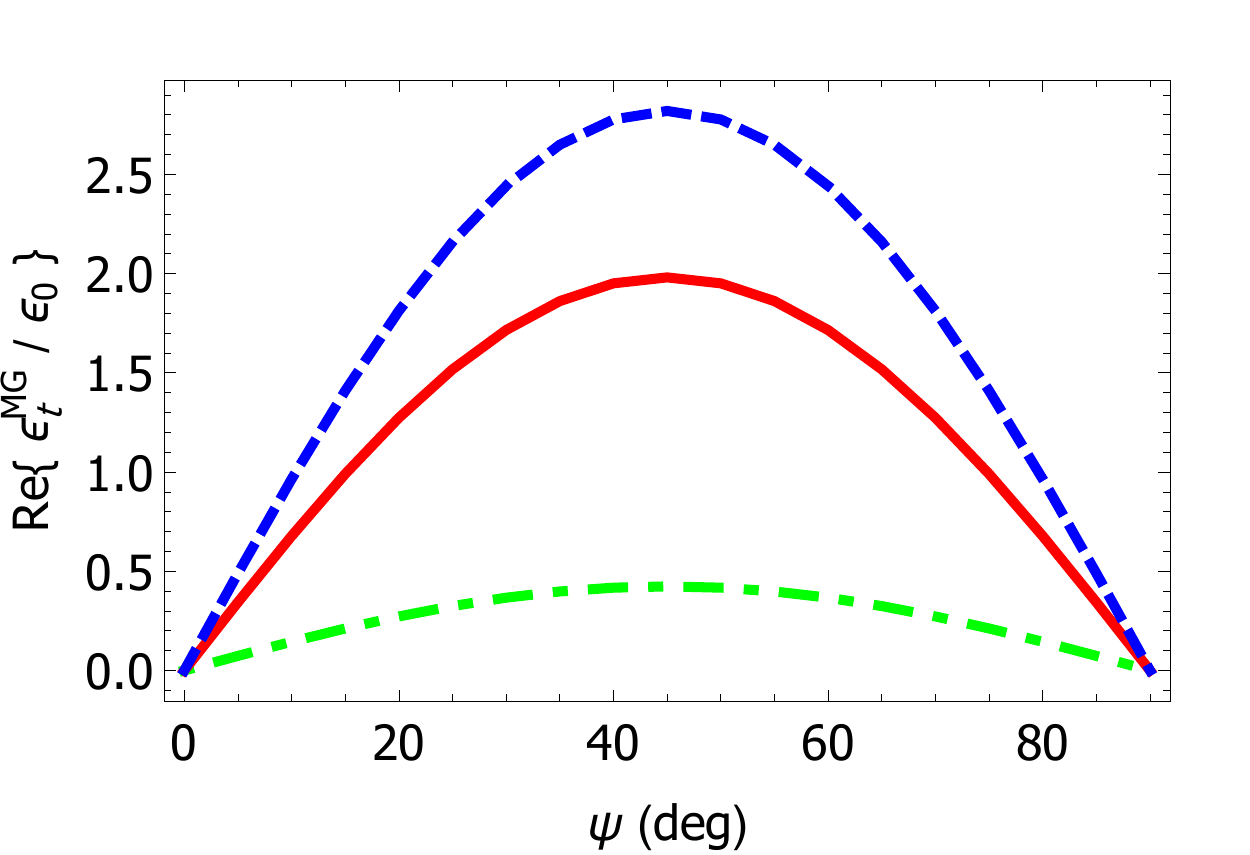} \hfill  \includegraphics[width=6.9cm]{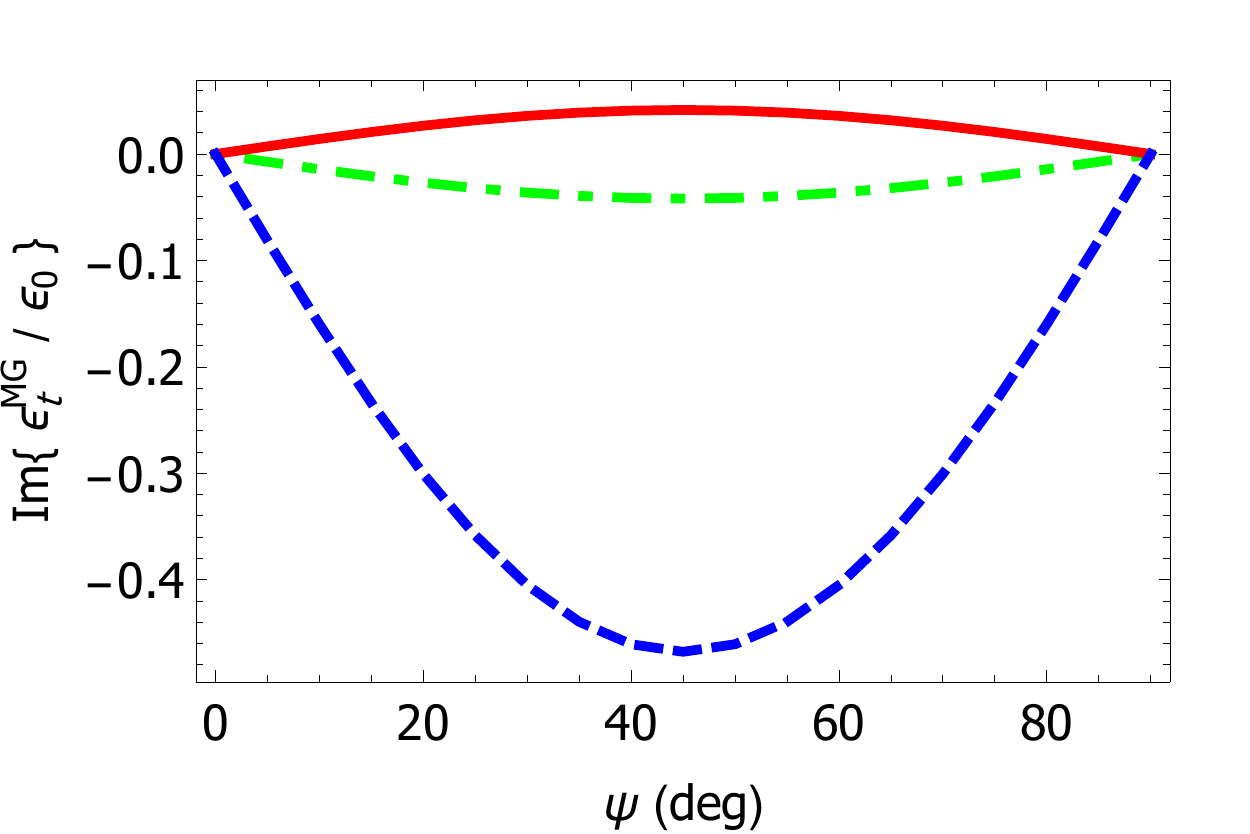} \\
 \caption{\label{Fig3} 
 Real and imaginary parts of  the non-zero components of the HCM's relative permittivity dyadic   
 versus orientation angle $\psi$ of gold nanoparticles for free-space wavelengths $\lambdao = 450$ nm  (green, broken dashed curve), 600 nm (red, solid curve), and 750 nm (blue, dashed curve). Size parameter $\rho = 0.2/\ko$, shape parameter $U=3$,  and volume fractions $f_a = f_b = 0.15$.
}
\end{figure}

Next we turn to the dependency on inclusion orientation, as specified by the angle $\psi\in[0,\pi/2]$.  The real and imaginary parts of $\eps^{MG}_{x,y,z,t}/\epso$ are plotted against $\psi$ in Fig.~\ref{Fig3} for $U=3$, $\rho = 0.2 / \ko$, and $f_a = f_b = 0.15$.
Both real and imaginary parts
 of $\eps^{MG}_{x,y,t} $
 are acutely sensitive to $\psi$ but
$\eps^{MG}_{z} $ is independent of $\psi$,  for all free-space wavelengths considered.
Unlike   $\eps^{MG}_{x,y} $,   $\eps^{MG}_{t} $
is symmetric with respect to reflection about $\psi = \pi/4$ and is null valued for $\psi\in\lec0,\pi/2\ric$.

\begin{figure}[!htb]
\centering
 \includegraphics[width=6.9cm]{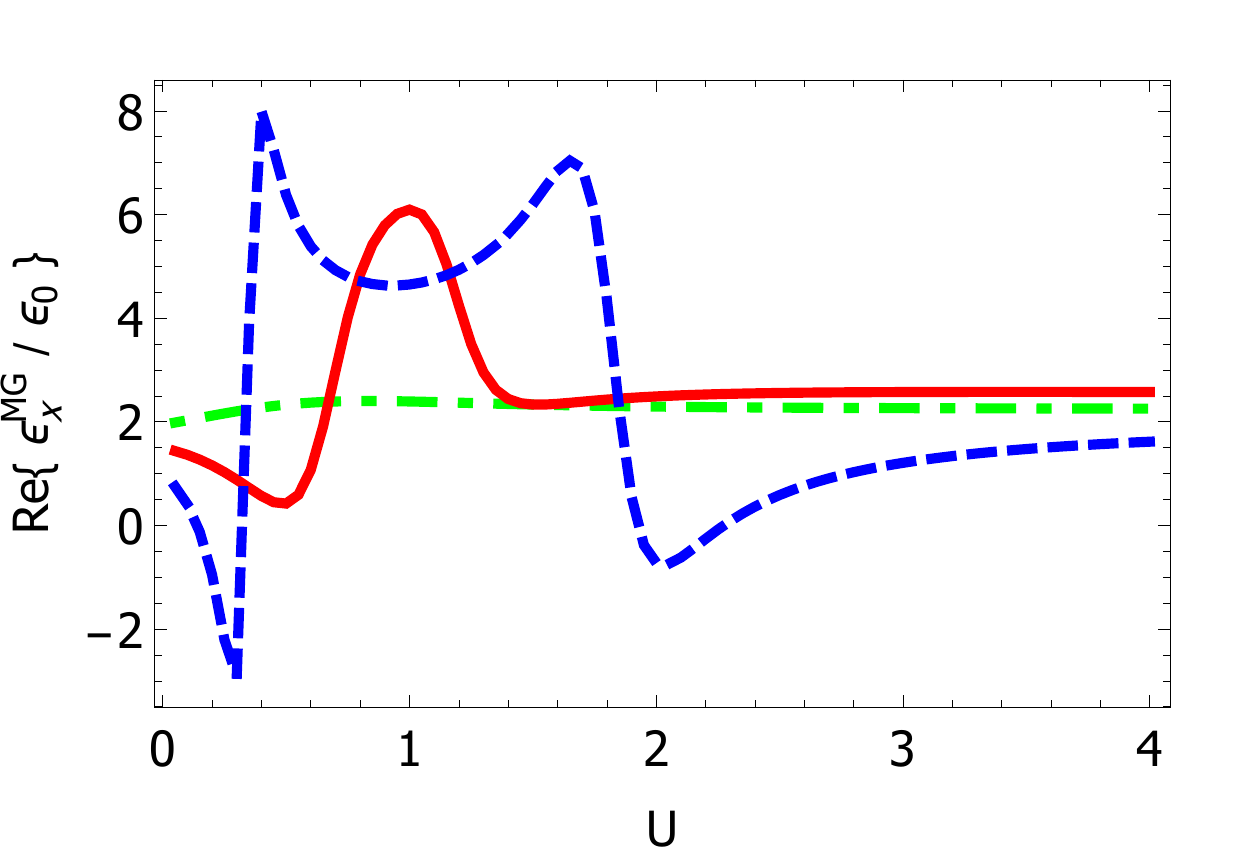} \hfill  \includegraphics[width=6.9cm]{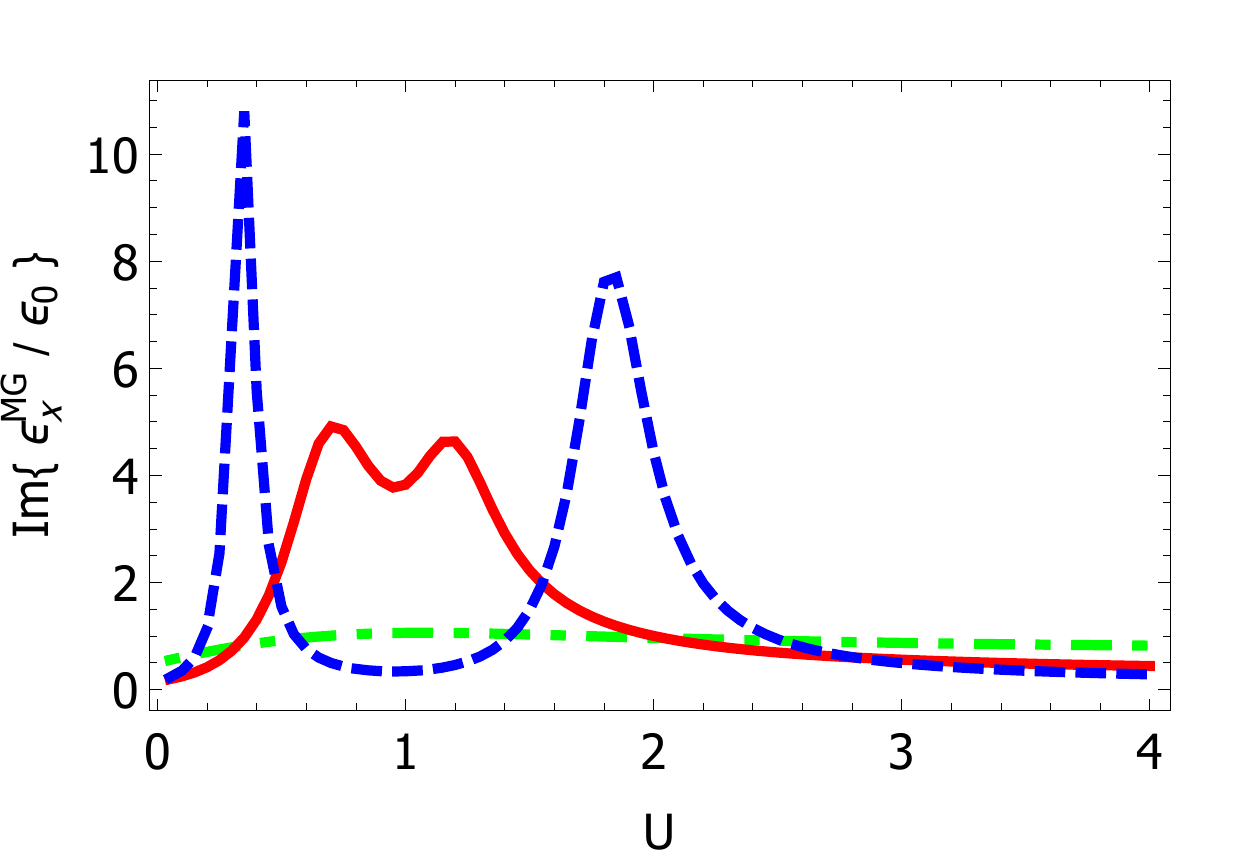} \\
 \includegraphics[width=6.9cm]{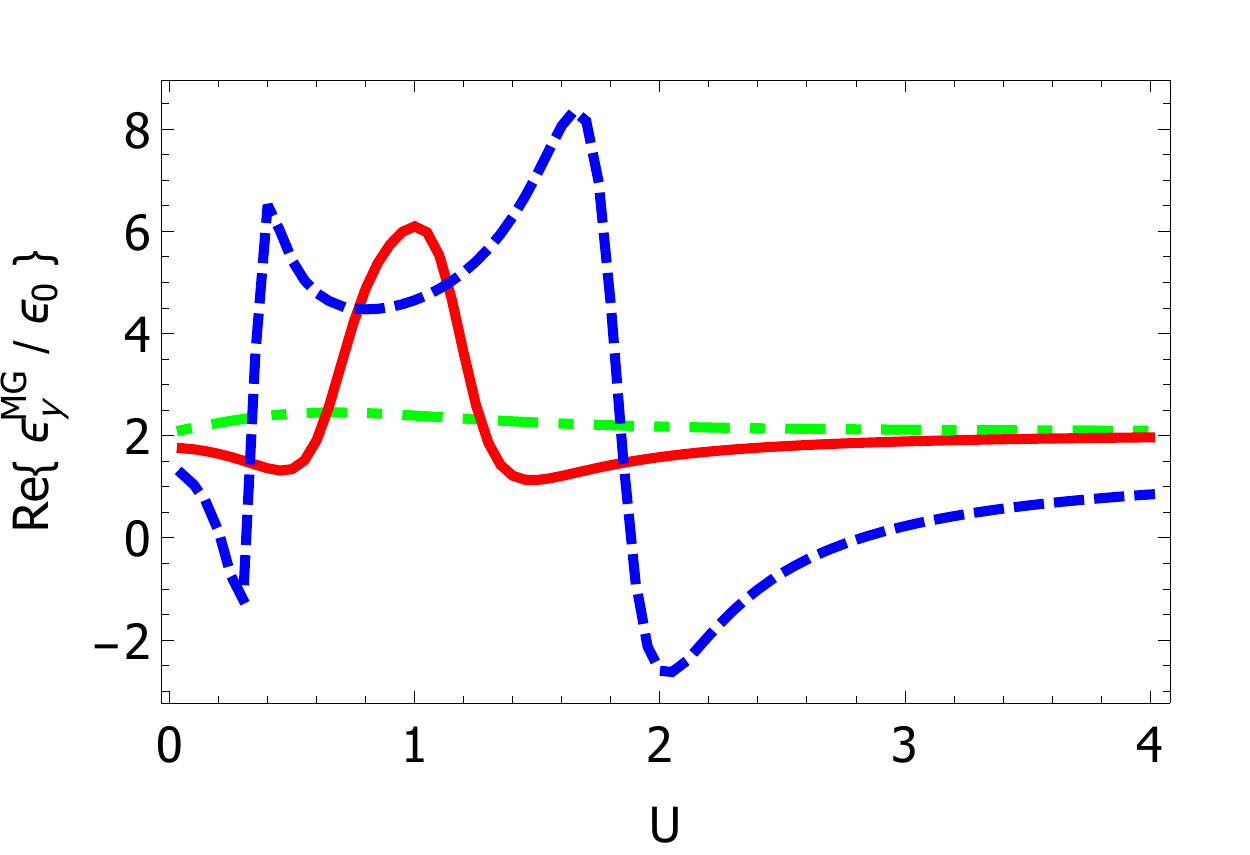} \hfill  \includegraphics[width=6.9cm]{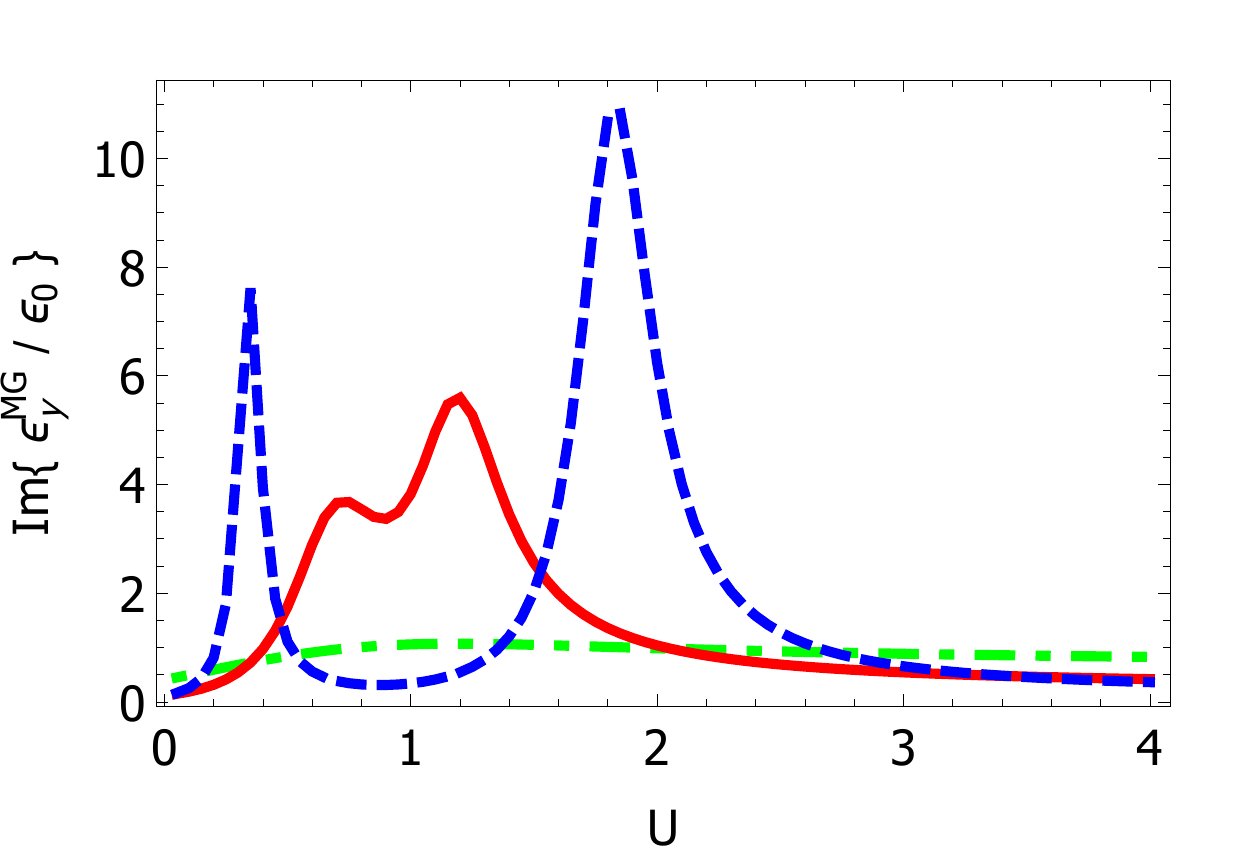} \\
 \includegraphics[width=6.9cm]{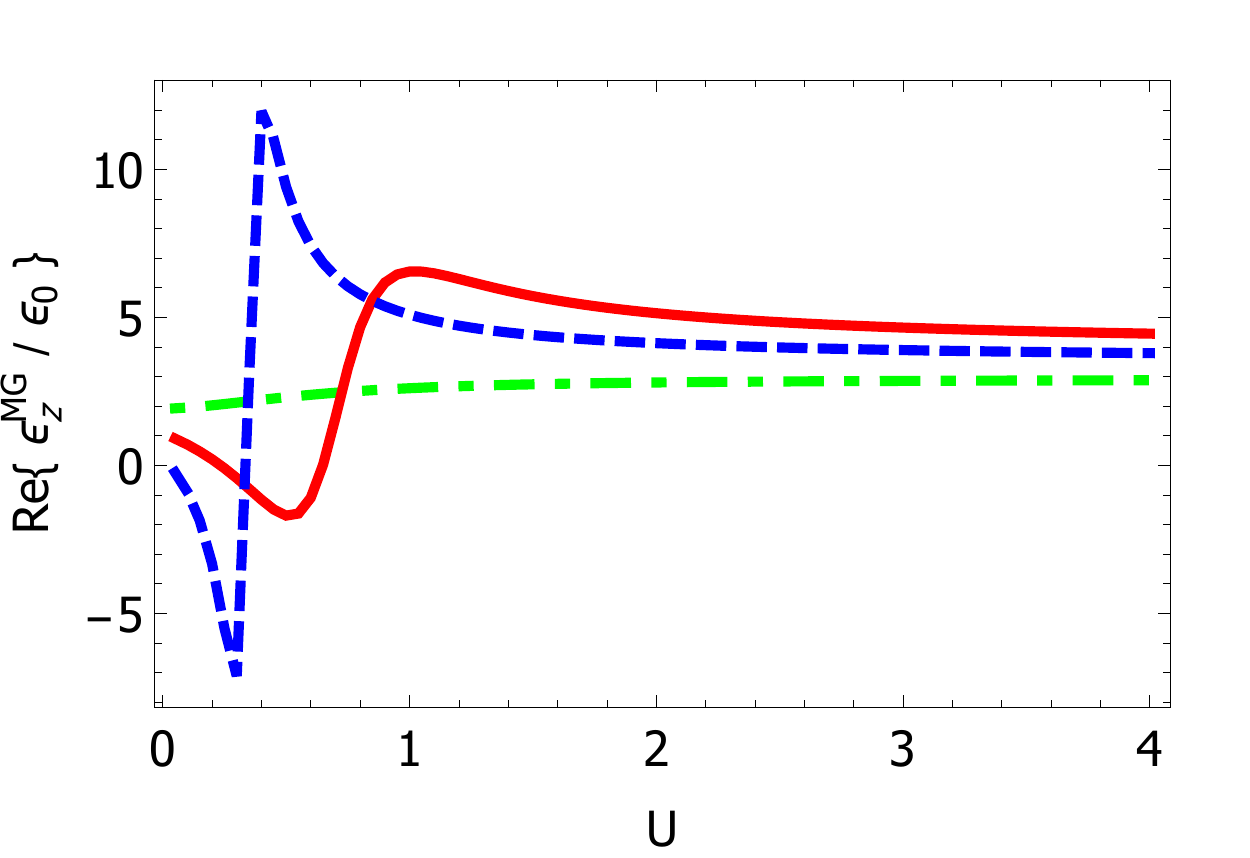} \hfill  \includegraphics[width=6.9cm]{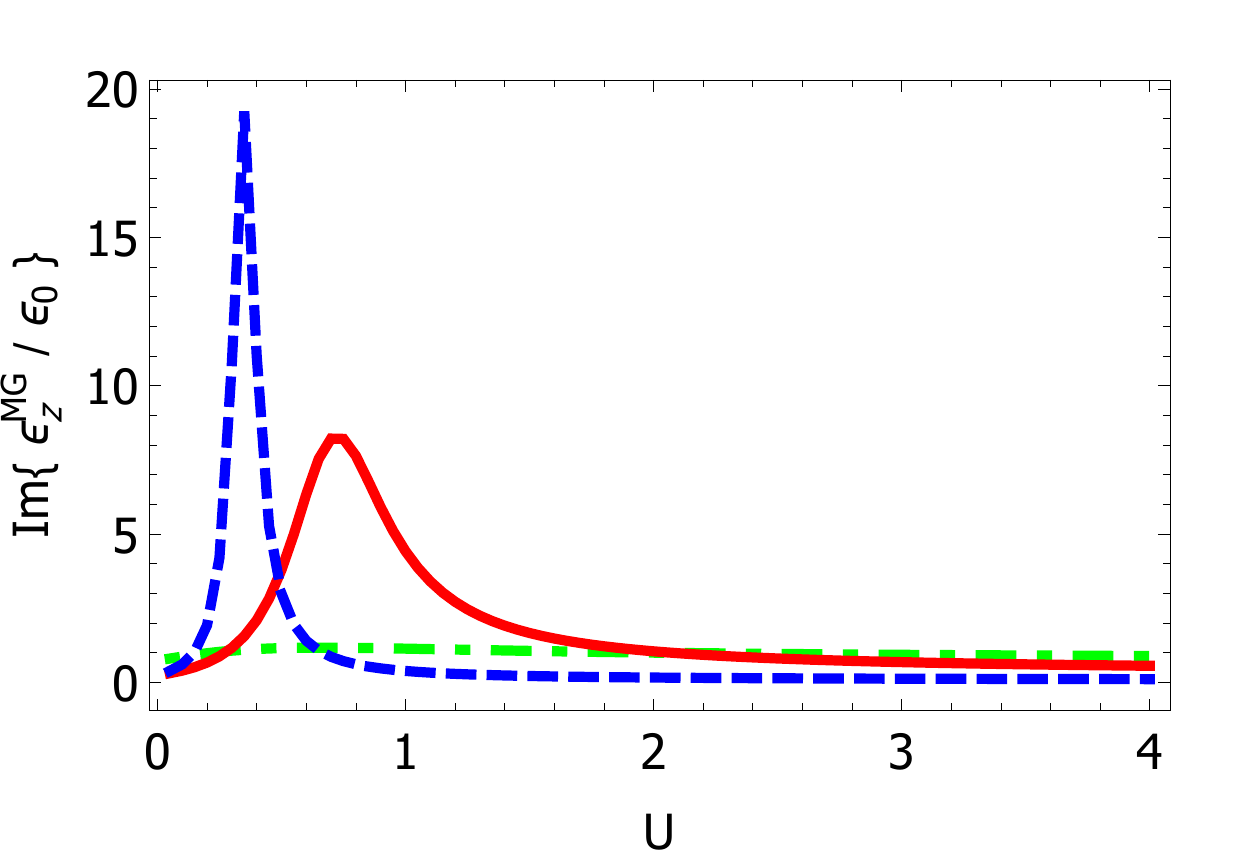} 
 \\
 \includegraphics[width=6.9cm]{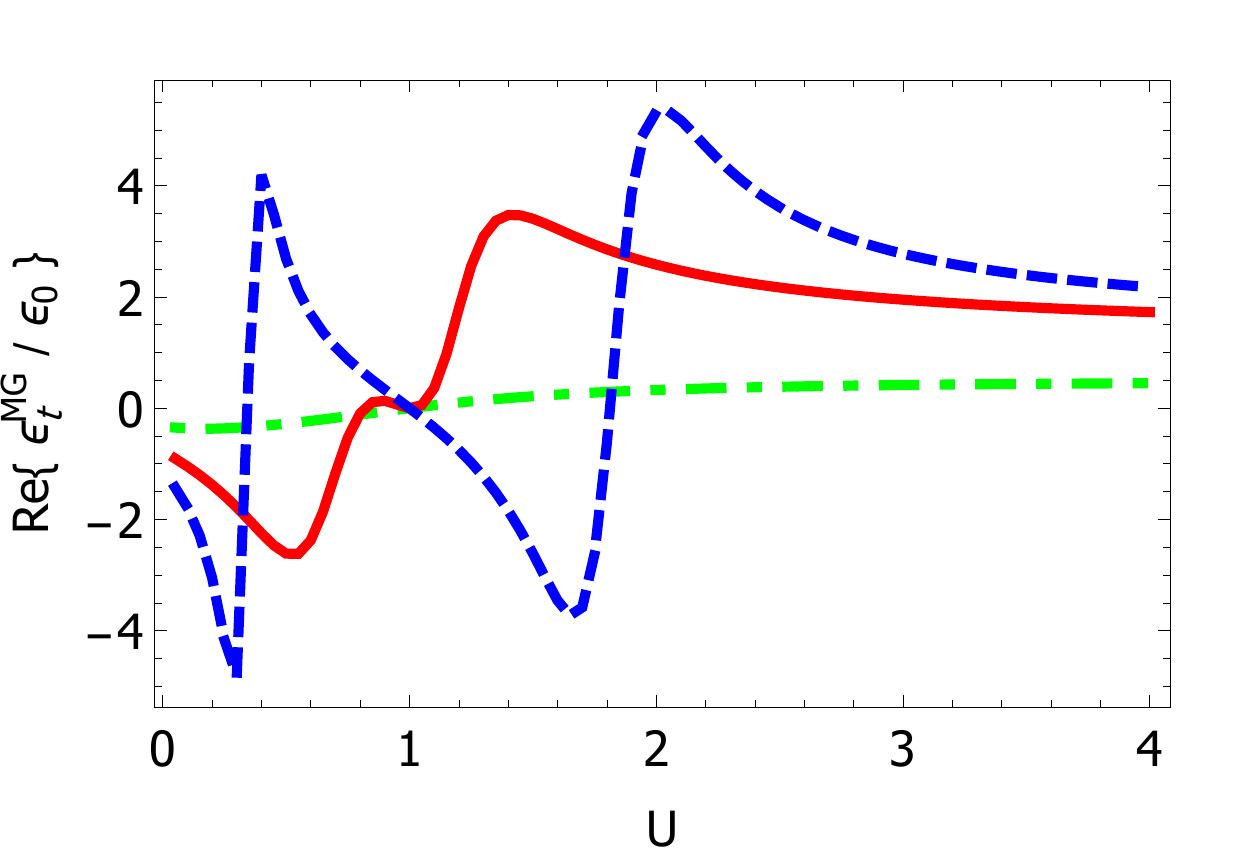} \hfill  \includegraphics[width=6.9cm]{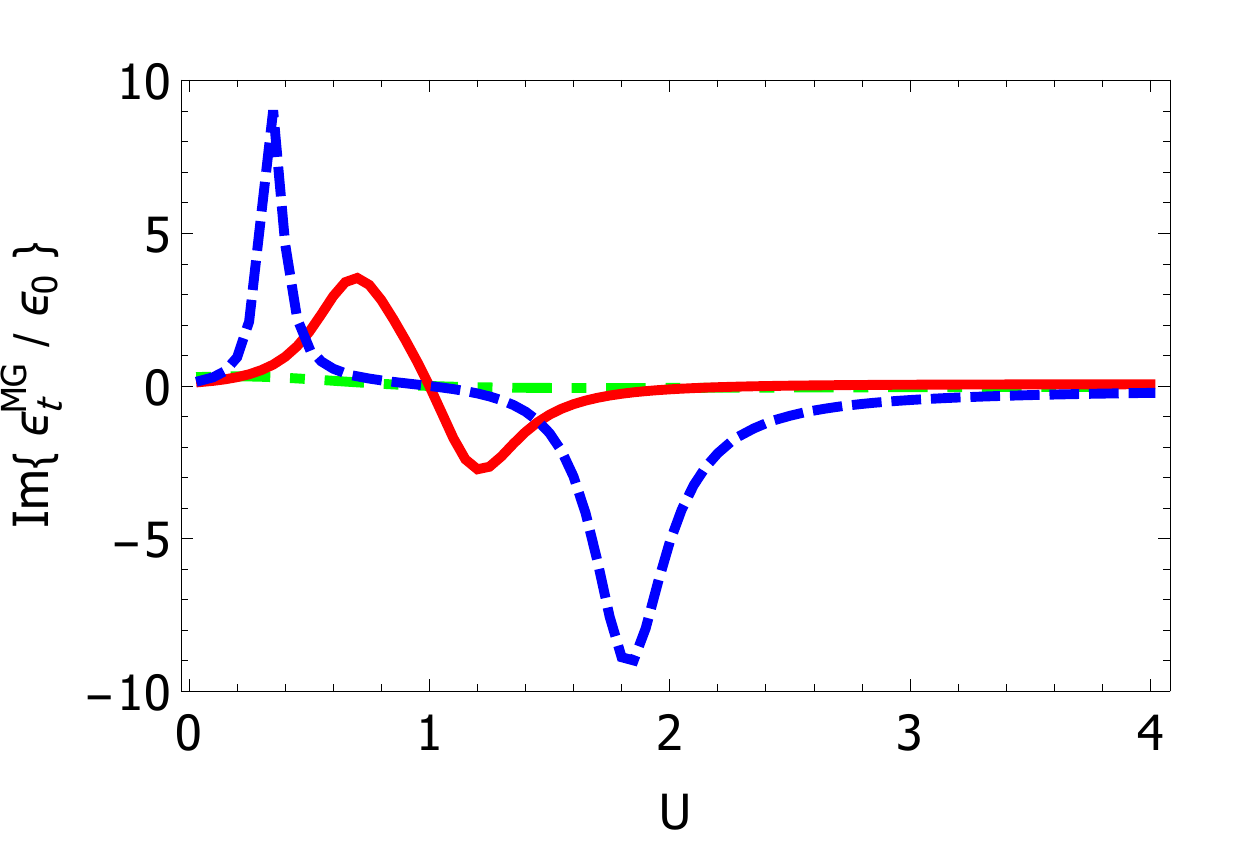} 
 \caption{\label{Fig4} 
  Real and imaginary parts of  the non-zero components of the HCM's relative permittivity dyadic   
 versus shape parameter $U$ of gold nanoparticles for free-space wavelengths $\lambdao = 450$ nm  (green, broken dashed curve), 600 nm (red, solid curve), and 750 nm (blue, dashed curve). Size parameter $\rho = 0.2/\ko$, orientation angle  $\psi=50^\circ$,  and volume fractions $f_a = f_b = 0.15$.
}
\end{figure}

Now the inclusion shape is considered via the shape parameter $U\in(0,4]$.  
 For $\rho = 0.2 / \ko$,   $\psi = 50^\circ$, and $f_a = f_b = 0.15$, 
 the real and imaginary parts of $\eps^{MG}_{x,y,z,t}/\epso$ are plotted against $U$ in Fig.~\ref{Fig4}.
Both the real and imaginary parts
 of $\eps^{MG}_{x,y,z,t}/\epso$
 are acutely sensitive to $U$ for $\lambdao \in\lec 600,750\ric$~nm
 but  both are quite insensitive to $U$  for $\lambdao = 450$ nm,  for $U<3$.
 For $U>3$, $\eps^{MG}_{x,y,z,t}/\epso $ is largely insensitive to $U$.
 Whereas the plots for
 $\eps^{MG}_{x}/\epso$ and $\eps^{MG}_{y}/\epso$ are qualitatively similar, those for
 $\eps^{MG}_{z}/\epso$ and $\eps^{MG}_{t}/\epso$ are somewhat different.

\begin{figure}[!htb]
\centering
 \includegraphics[width=6.9cm]{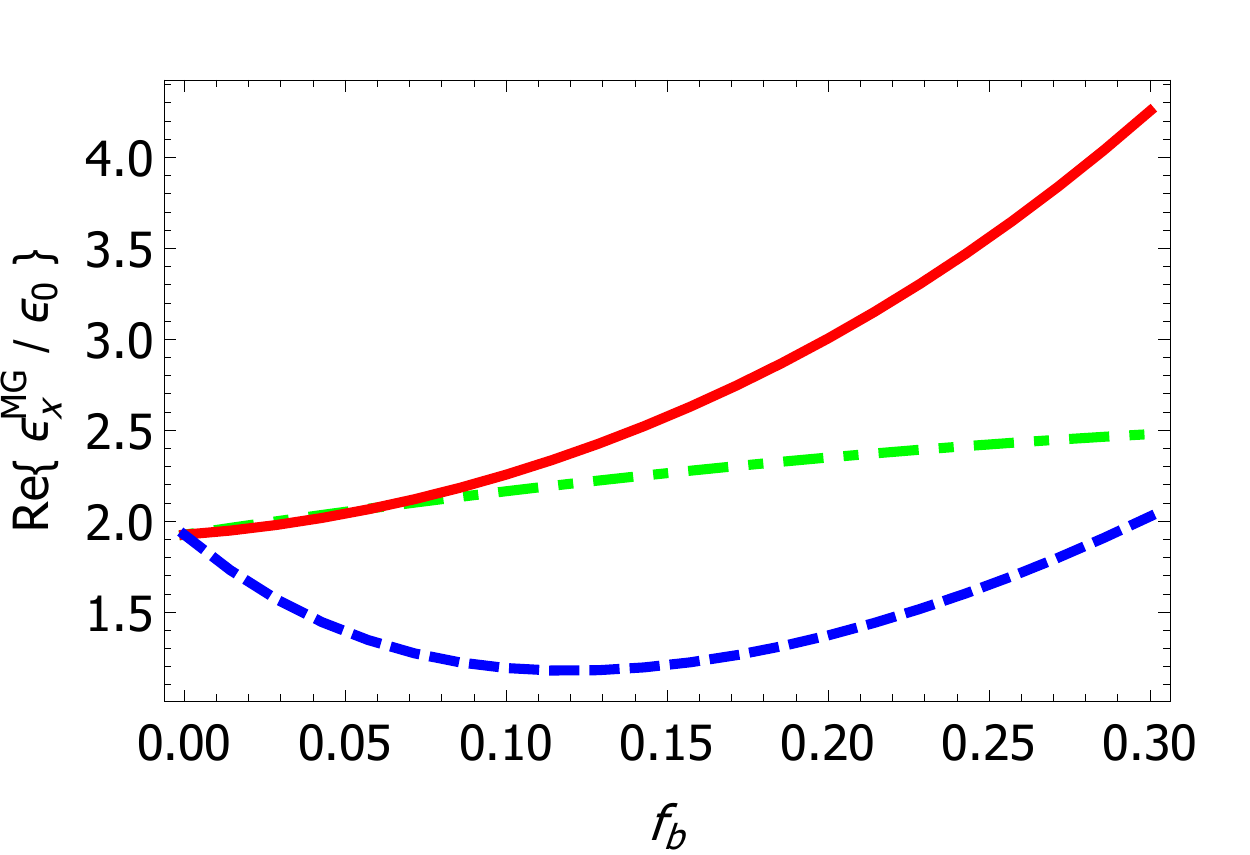} \hfill  \includegraphics[width=6.9cm]{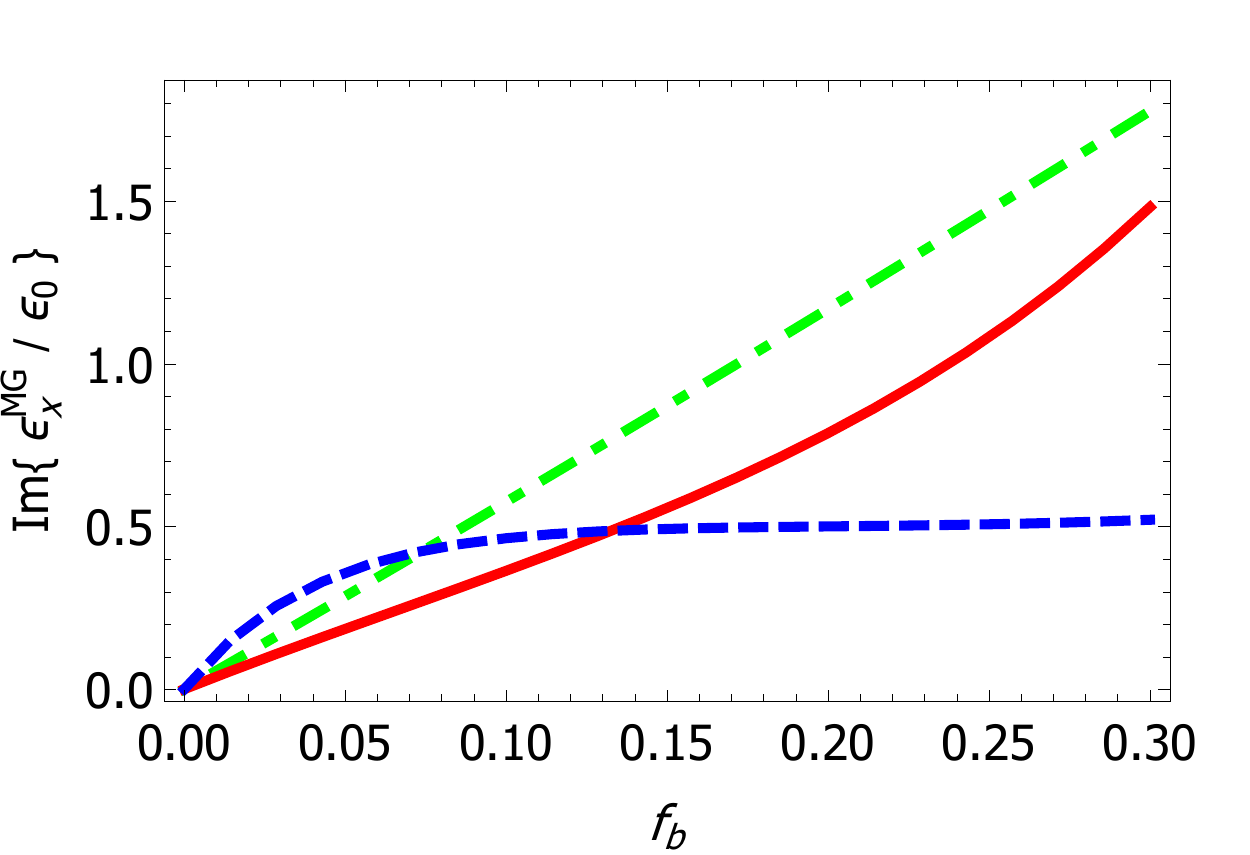} \\
 \includegraphics[width=6.9cm]{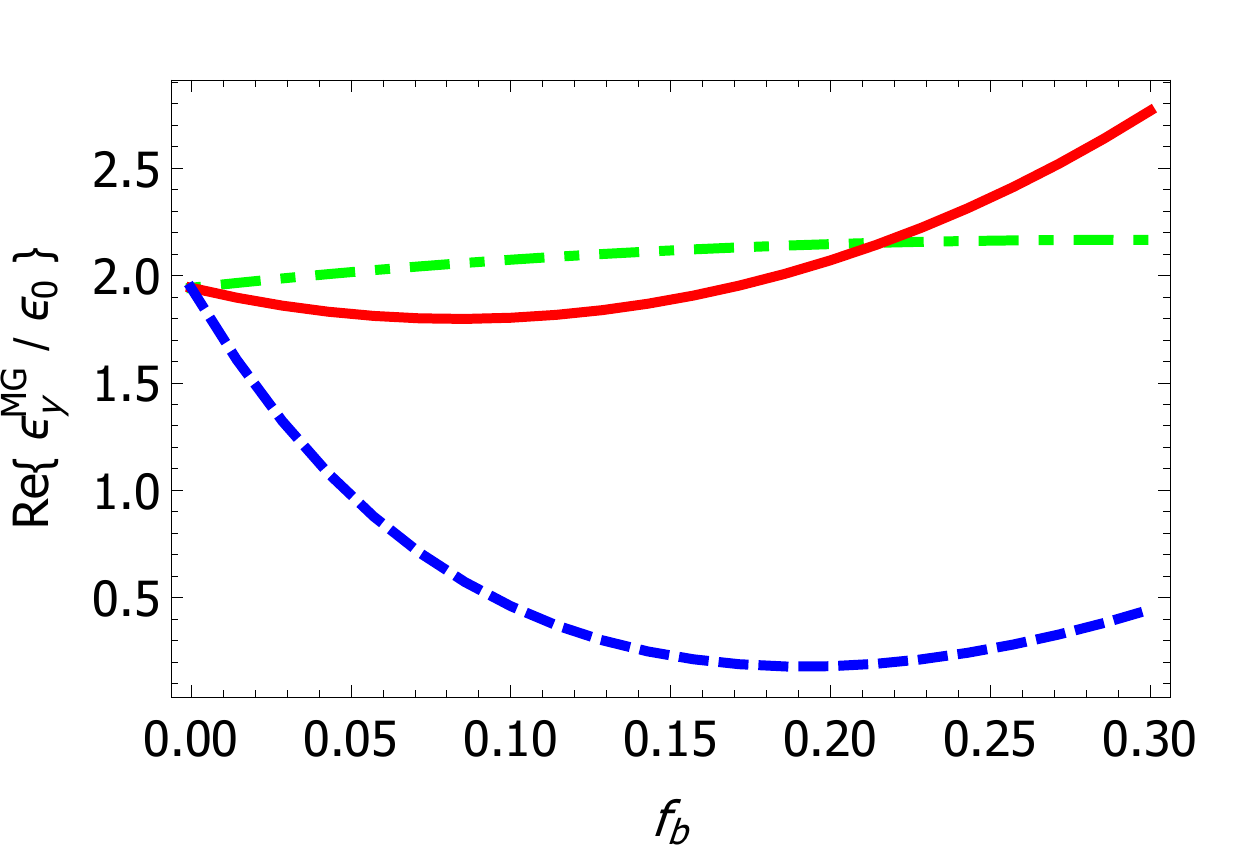} \hfill  \includegraphics[width=6.9cm]{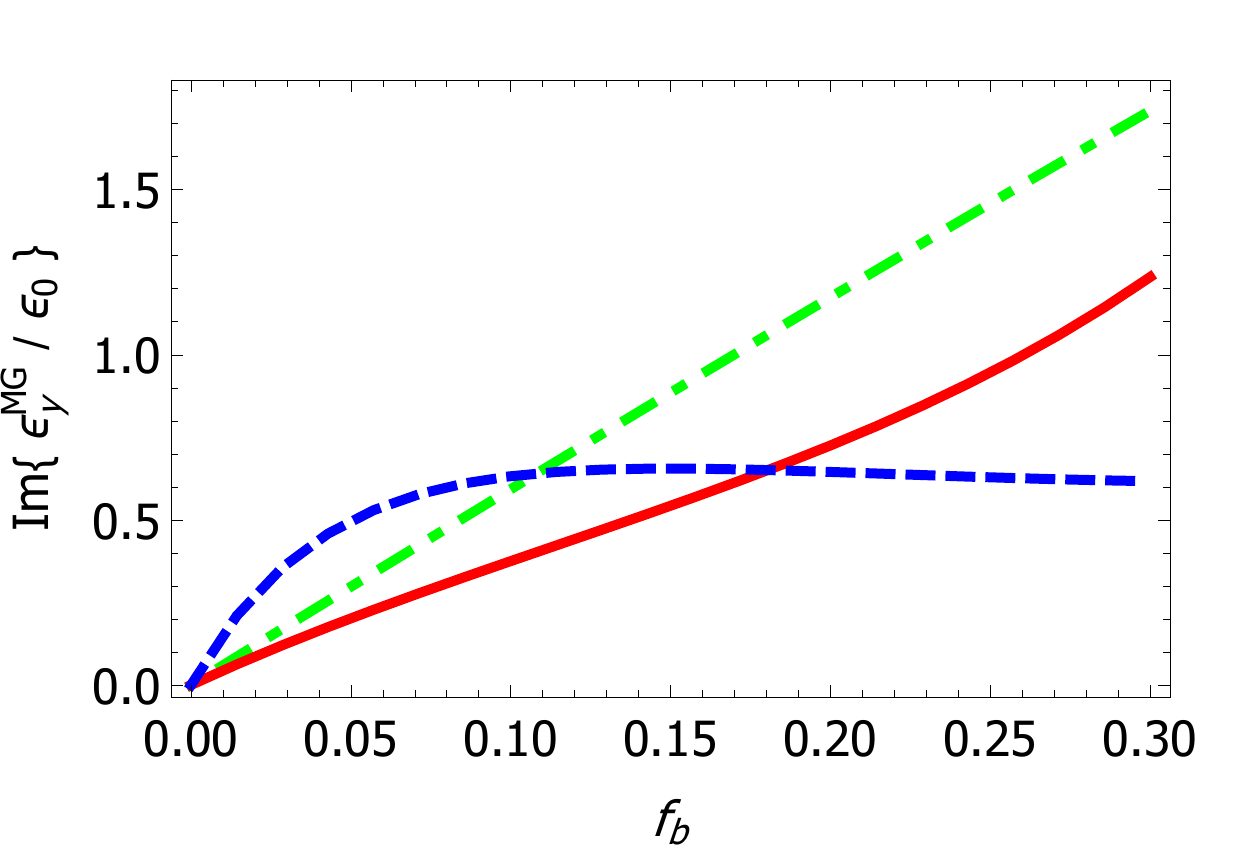} \\
 \includegraphics[width=6.9cm]{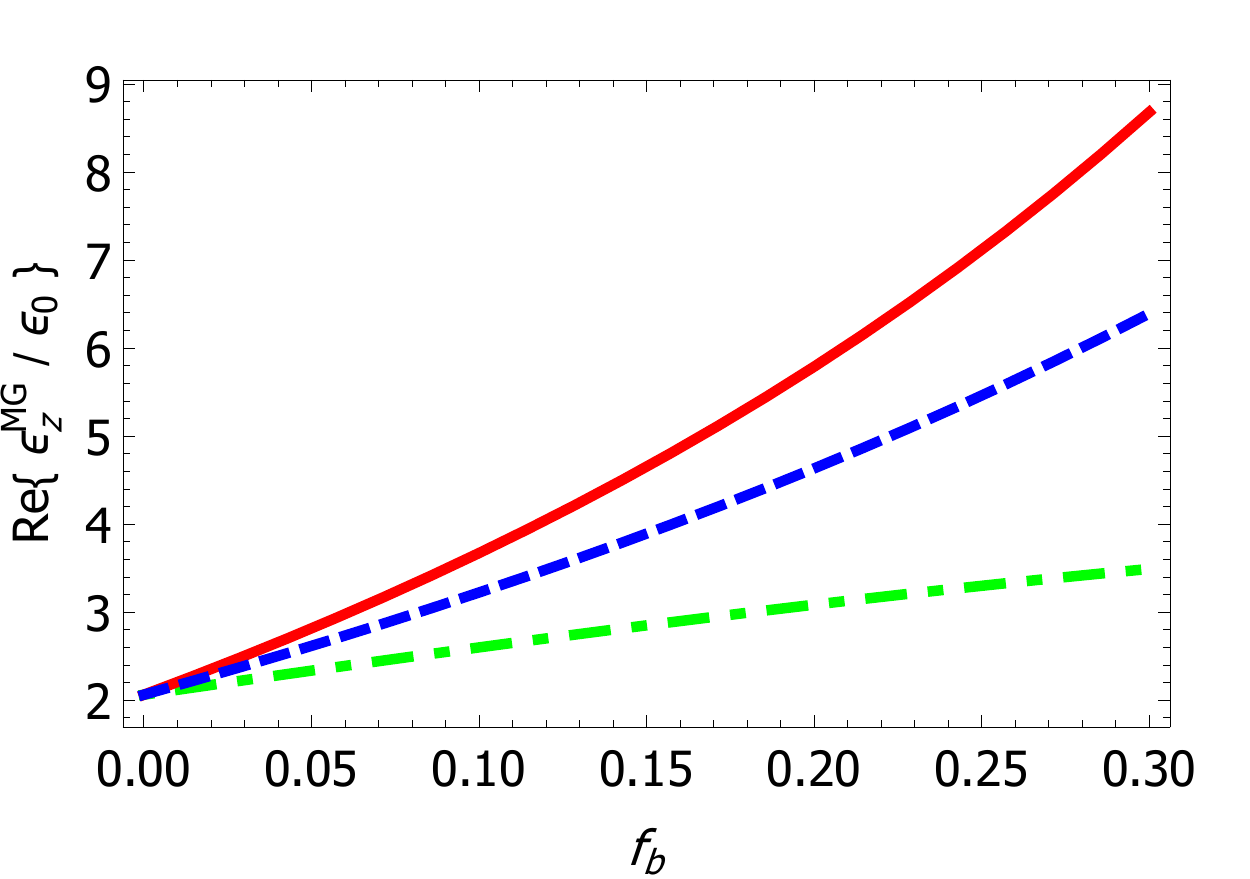} \hfill  \includegraphics[width=6.9cm]{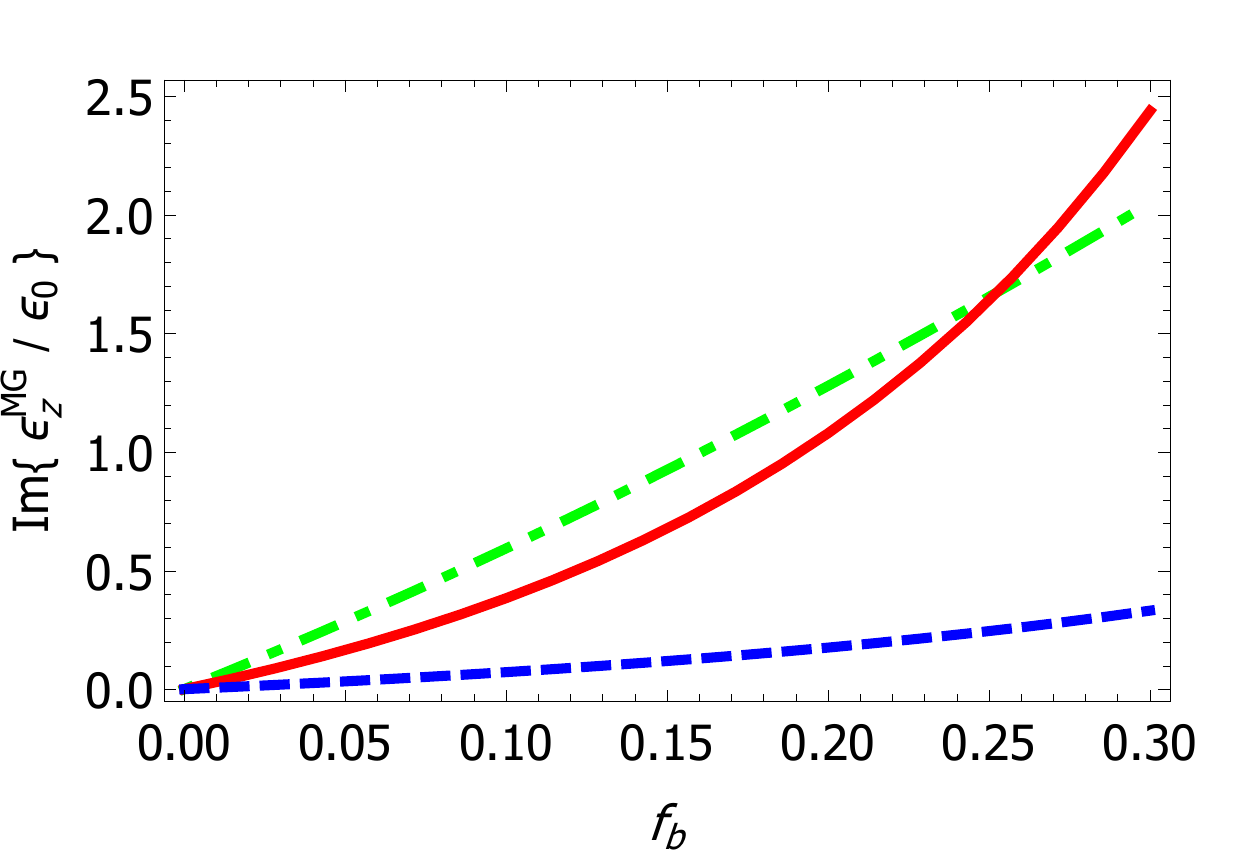} \\
 \includegraphics[width=6.9cm]{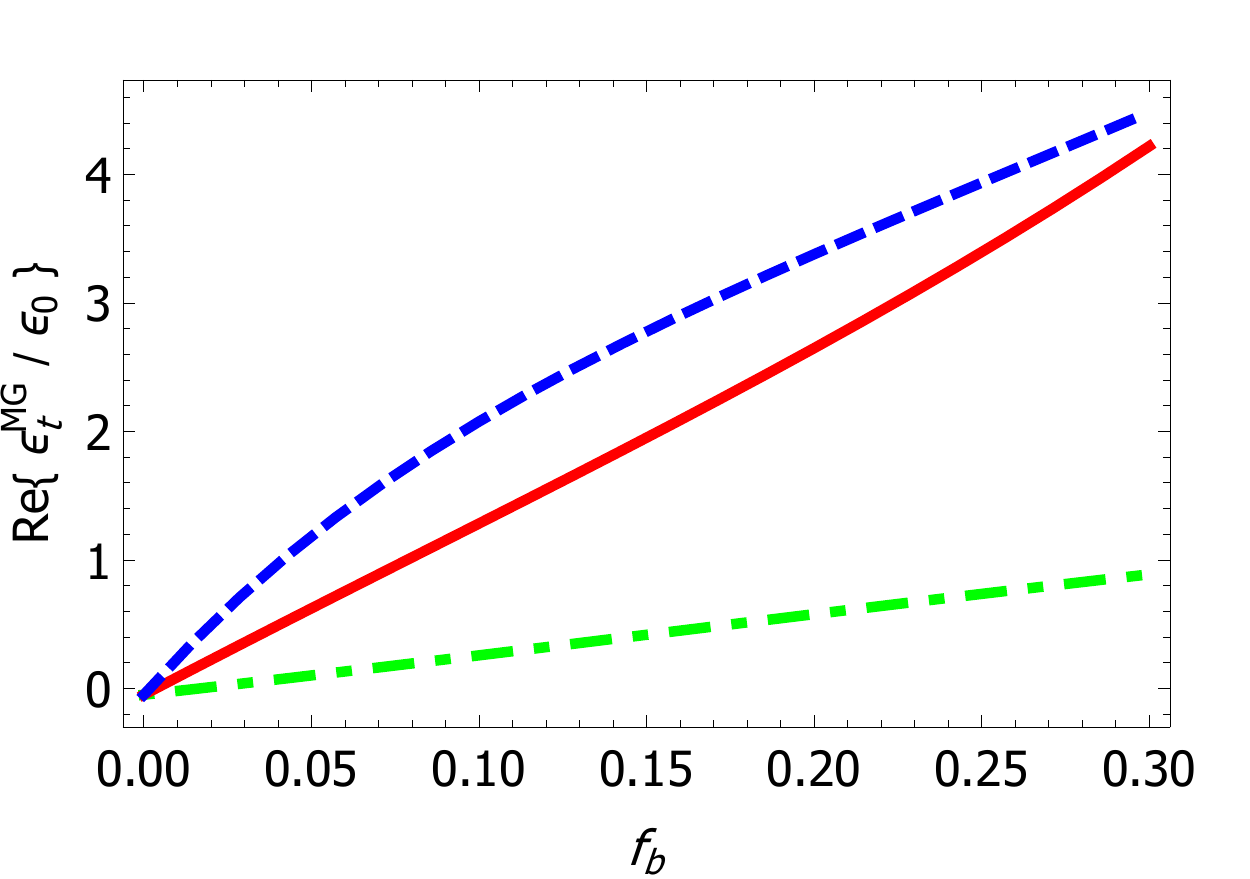} \hfill  \includegraphics[width=6.9cm]{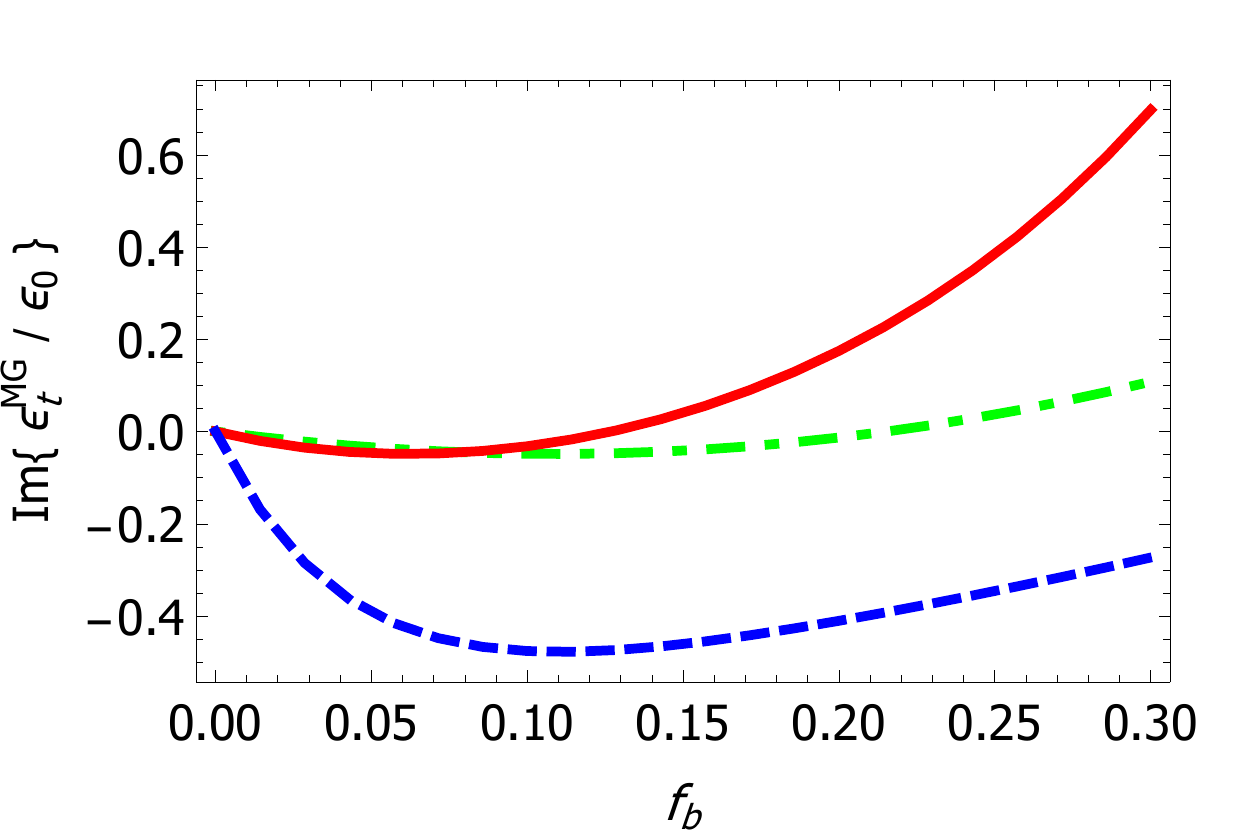} 
 \caption{\label{Fig5} 
  Real and imaginary parts of  the non-zero components of the HCM's relative permittivity dyadic   
 versus volume fraction of gold   for free-space wavelengths $\lambdao = 450$ nm  (green, broken dashed curve), 600 nm (red, solid curve), and 750 nm (blue, dashed curve). Size parameter $\rho = 0.2/\ko$, shape parameter $U=3$, orientation angle $\psi=50^\circ$,  and volume fraction $f_a =  0.3 - f_b$.
}
\end{figure}

The sensitivity of the HCM's permittivity dyadic  to the volume fraction of the inclusions is explored next.
In Fig.~\ref{Fig5}, the real and imaginary parts of $\eps^{MG}_{x,y,z,t}/\epso$ are plotted against $f_b$, with  $\rho = 0.2 / \ko$,   $U=3$, 
$\psi = 50^\circ$, and $f_a = 0.3 - f_b$. 
As in Figs.~\ref{Fig2} and \ref{Fig4}, the plots of both the real and imaginary parts of $\eps^{MG}_{x}/\epso$ and $\eps^{MG}_{y}/\epso$ in Fig.~\ref{Fig5} are qualitatively similar for all wavelengths considered. The real parts of  $\eps^{MG}_{z}/\epso$ and $\eps^{MG}_{t}/\epso$, as well as the imaginary part of $\eps^{MG}_{z}/\epso$,  increase uniformly as $f_b$ increases, whereas the imaginary part of $\eps^{MG}_{t}/\epso$ does not, for all wavelengths considered.
  
  The  numerical results presented in Figs.~\ref{Fig2}--\ref{Fig5} were computed using the extended Maxwell Garnett formalism, per Eq.~\r{epsMG}, incorporating the size-dependent permittivity of gold nanoparticles delivered by Eqs.~\r{Au} and \r{Au2}. Commonly, the unextended version of the Maxwell Garnett formalism \c{MAEH}~---~which arises from the extended version in the limit $\rho \to 0$~---~is implemented. 
  It is also common for size-independent dielectric properties 
of the inclusion materials to be incorporated in the homogenization formalism used. Indeed, in a recent numerical study of gold-impregnated vaterite \c{Noskov}, a size-independent permittivity scalar was adopted for gold nanoparticles  along with a non-rigorous version of the Maxwell Garnett formalism that fails to properly account for the uniaxial anisotropy of 
  vaterite in the depolarization dyadics. In the case of spherical inclusions of air and gold (i.e., $U=1$), this 
  non-rigorous
  formalism's  estimate of the HCM's  permittivity dyadic 
  is given by
  \begin{eqnarray}
  \l{Noskov_eq}
  \={\tilde{\eps}}_{\,MG} &=& \=\eps_{\,c}
  + \Bigg\{ f_a \le \eps_a \=I - \=\eps_{\,c} \ri
  \nonumber \\ && 
   \. \les  
  \=\eps_{\,c} + \frac{1-f_a}{3} \le \eps_a \=I - \=\eps_{\,c} \ri
  \ris^{-1}  \nonumber \\ && 
  +
  f_b \le \eps_b \=I - \=\eps_{\,c} \ri
  \nonumber \\ && 
   \. \les  
  \=\eps_{\,c} + \frac{1-f_b}{3} \le \eps_b \=I - \=\eps_{\,c} \ri
  \ris^{-1}
  \Bigg\}
   \. \=\eps_{\,c}.
  \end{eqnarray}

\begin{figure}[!htb]
\centering
 \includegraphics[width=6.9cm]{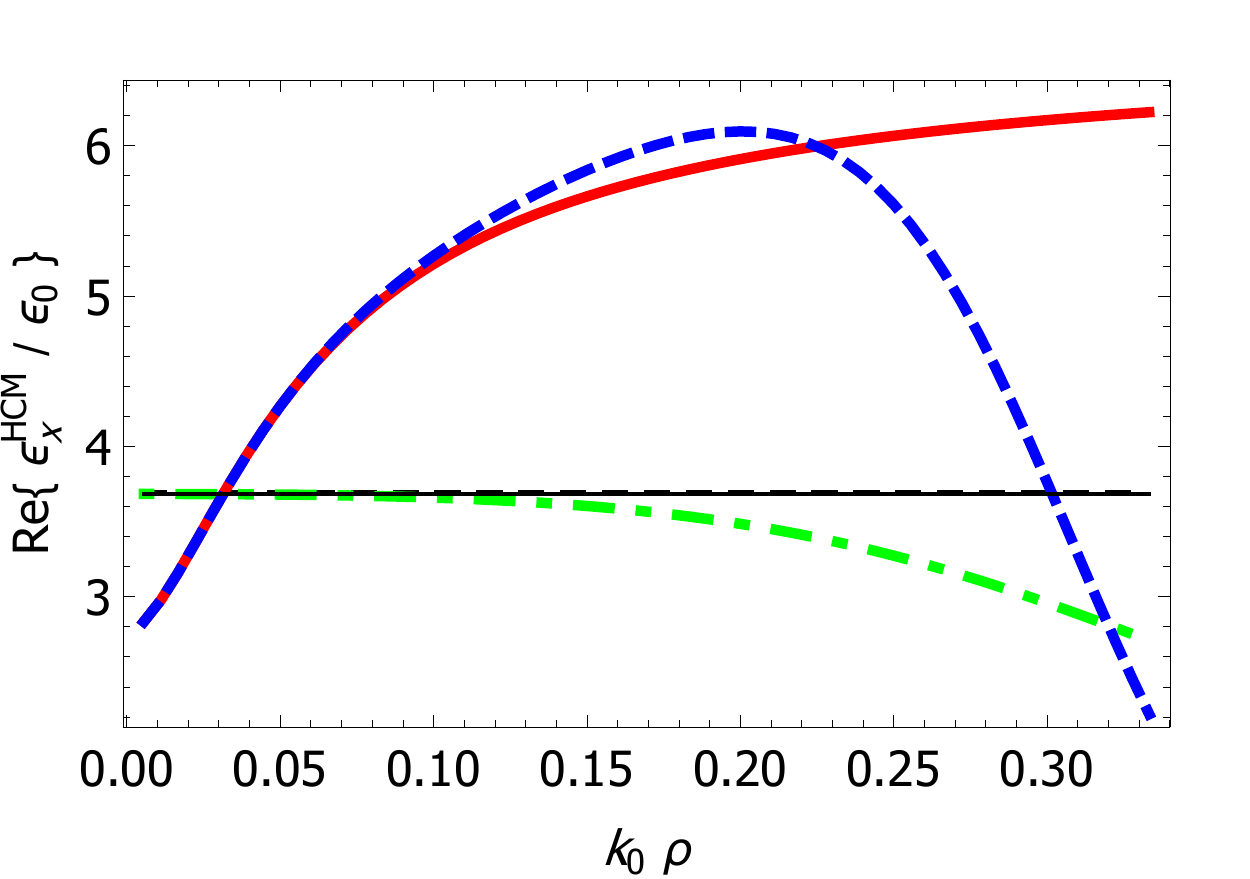} \hfill  \includegraphics[width=6.9cm]{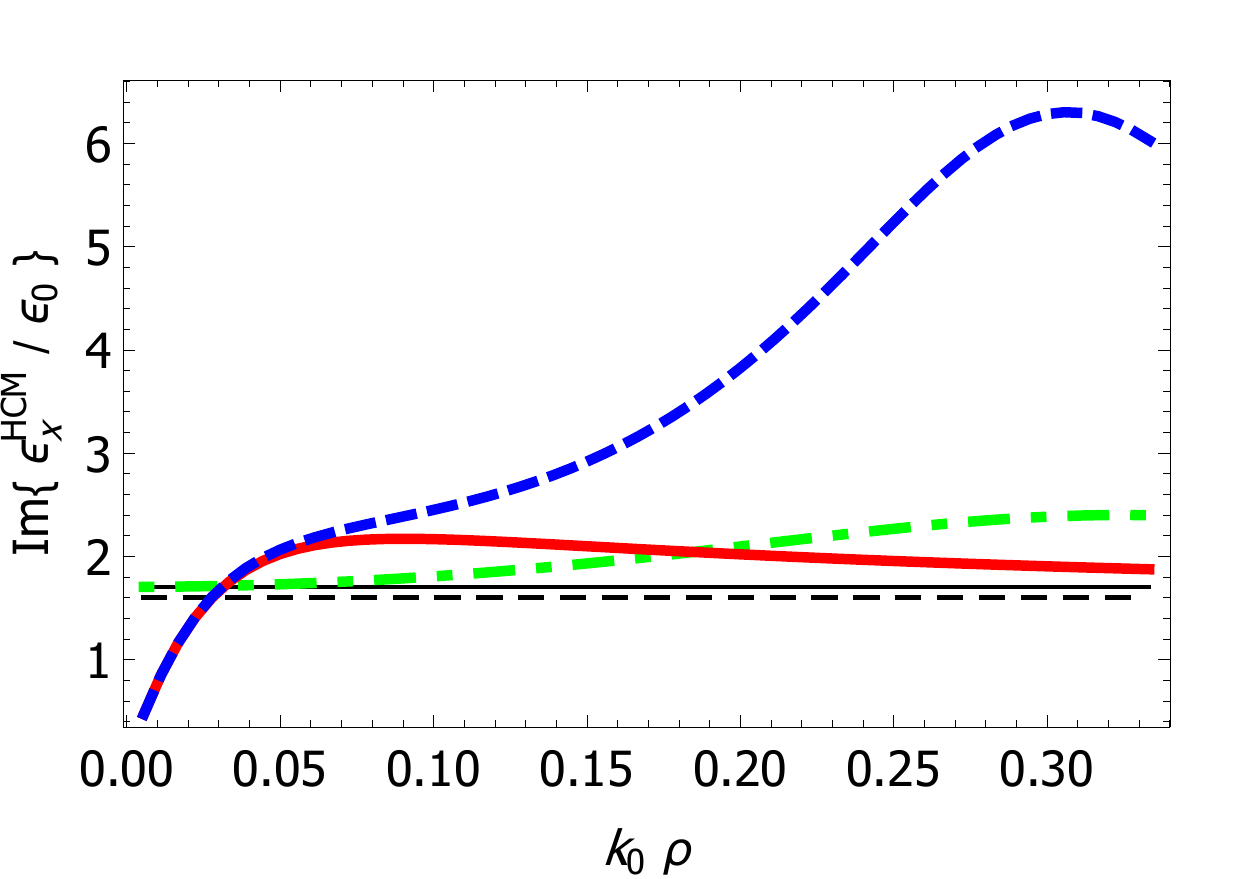} \\
 \includegraphics[width=6.9cm]{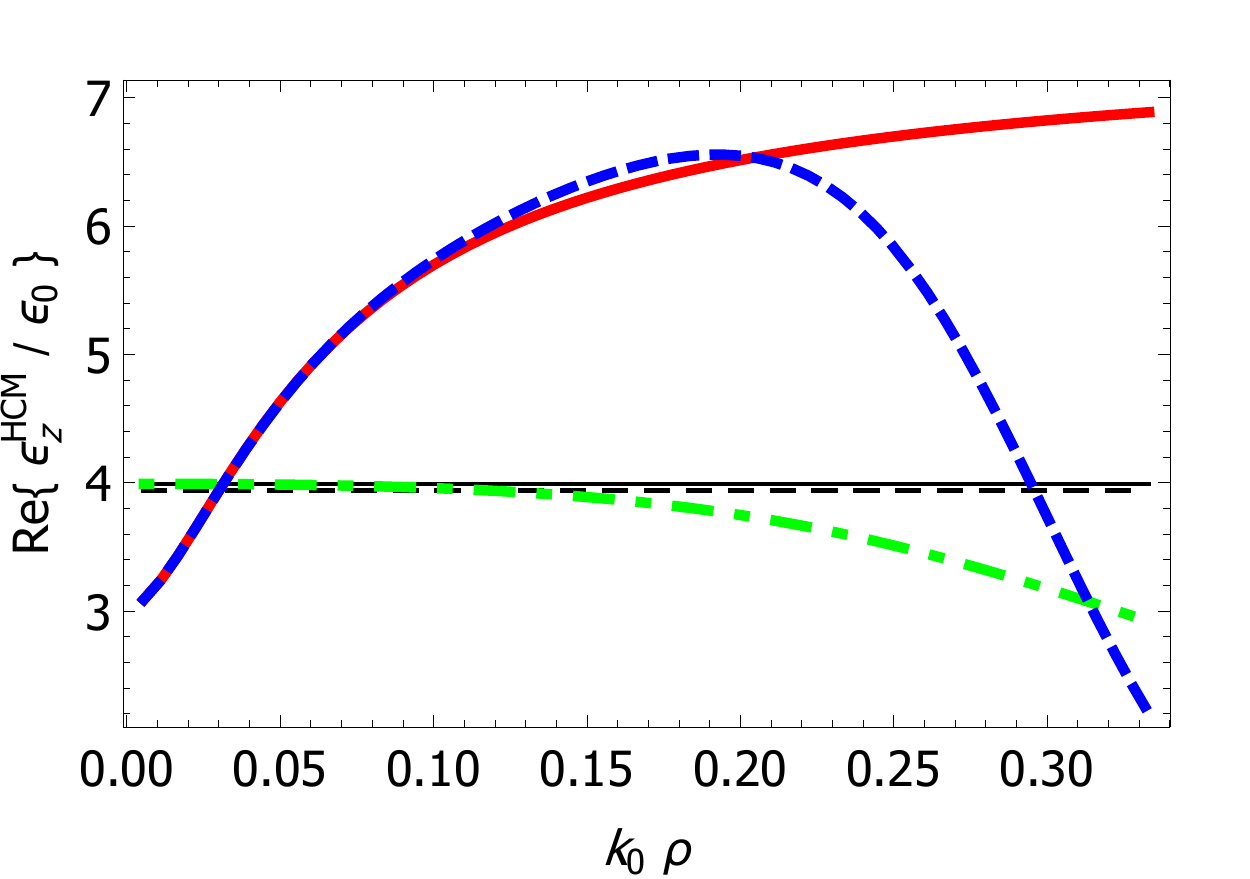} \hfill  \includegraphics[width=6.9cm]{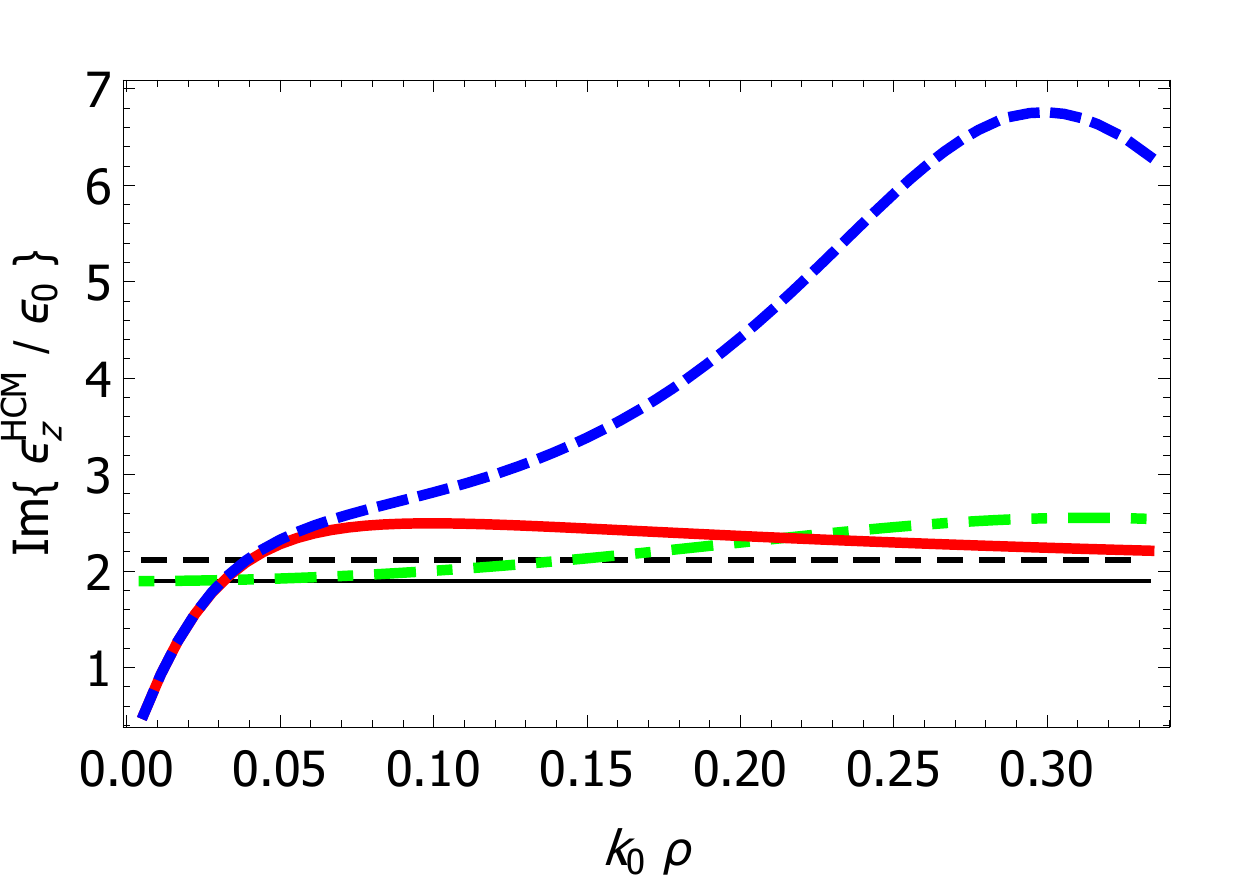} 
 \caption{\label{Fig6} 
 Real and imaginary parts of  the non-zero components of the HCM's relative permittivity dyadic   
 versus relative size of gold nanoparticles. (a)
 Blue, dashed curves: estimates delivered by the extended Maxwell Garnett formalism, with size-dependent
  $\eps_b$. (b) Green, broken-dashed curves: as (a) but computed using size-independent $\eps_b = \le -7.1698+ i 6.5253 \ri \epso$.
  (c) Red, solid curves: as (a) but computed using unextended  Maxwell Garnett formalism. (d) Black, solid horizontal  lines: estimates delivered by the unextended Maxwell Garnett formalism, with size-independent  $\eps_b = \le -7.1698+ i 6.5253 \ri \epso$.
  (e) Black,  dashed horizontal lines: estimates delivered by Eq.~\r{Noskov_eq}, with size-independent  $\eps_b = \le -7.1698+ i 6.5253 \ri \epso$.
 Free-space wavelength $\lambdao =  600$ nm,
  shape parameter $U=1$, and volume fractions $f_a = f_b = 0.15$.
}
\end{figure}

  It is of interest to compare and contrast the estimate of $\=\eps_{\,HCM}$ provided by the  extended Maxwell Garnett formalism, incorporating the size-dependent permittivity of gold nanoparticles, with estimates provided by these related approaches in which the sizes of the inclusions  and/or
  the anisotropy of vaterite are not taken into account. For this purpose, let us fix the free-space wavelength $\lambdao =  600$ nm,
  shape parameter $U=1$ 
  (in which case the orientation angle $\psi$ becomes irrelevant), and volume fractions $f_a = f_b = 0.15$. Since the inclusions are spherical, the HCM's permittivity dyadic has the   uniaxial form
  \begin{equation} \l{HCM_uniaxial}
  \=\eps_{\,HCM}  = \eps^{HCM}_x \le \ux\,\ux
   +  \uy\,\uy \ri + \eps^{HCM}_z \uz\,\uz.
\end{equation}
  In Fig.~\ref{Fig6}, the real and imaginary parts of $\eps^{HCM}_x/\epso$ and $\eps^{HCM}_z/\epso$
   are plotted against $\ko \rho$ using  estimates  from
 \begin{itemize}
\item[(a)] $\=\eps^{MG}$
  computed using Eq.~\r{epsMG}
   with a size-dependent $\eps_b$ (blue, dashed curves);
 \item[(b)] $\=\eps^{MG}$
   as computed in (a)  but  with size-independent $\eps_b = \le -7.1698+ i 6.5253 \ri \epso$, which is the permittivity of gold nanoparticles of radius $\rho = 3$ nm per Eqs.~\r{Au} and \r{Au2} 
   (green, broken-dashed curves);
 \item[(c)]
   $\=\eps^{MG}$
   as computed in (a)  but 
   with the unextended  Maxwell Garnett formalism (red, solid curves); 
   \item[(d)] 
   $\=\eps^{MG}$ delivered by the unextended Maxwell Garnett formalism, with size-independent  $\eps_b = \le -7.1698+ i 6.5253 \ri \epso$
   (black, solid horizontal  lines); and
  \item[(e)] $  \={\tilde{\eps}}_{\,MG}$ delivered by Eq.~\r{Noskov_eq},
    with size-independent  $\eps_b = \le -7.1698+ i 6.5253 \ri \epso$ (black,  dashed horizontal lines). 
 \end{itemize}
All five estimates of the real and imaginary parts of $ \eps^{HCM}_x$ and $ \eps^{HCM}_z$ approximately agree at $\rho = 3$ nm. For smaller values of $\rho$,
the three estimates based on a size-independent $\eps_b$, i.e., (a), (d), and (e), approximately agree, and the two estimates based on a size-dependent $\eps_b$, i.e., (b) and (c), are in close agreement, but the estimates (b) and (c)   differ markedly from the estimates (a), (d) and (e). For larger values of $\rho$, especially for $\rho > 10$ nm, quite large differences emerge between the estimates (a), (b), and (c);  however, the estimates (d) and (e)   remain quite close to each other.
  
\begin{figure}[!htb]
\centering
 \includegraphics[width=6.9cm]{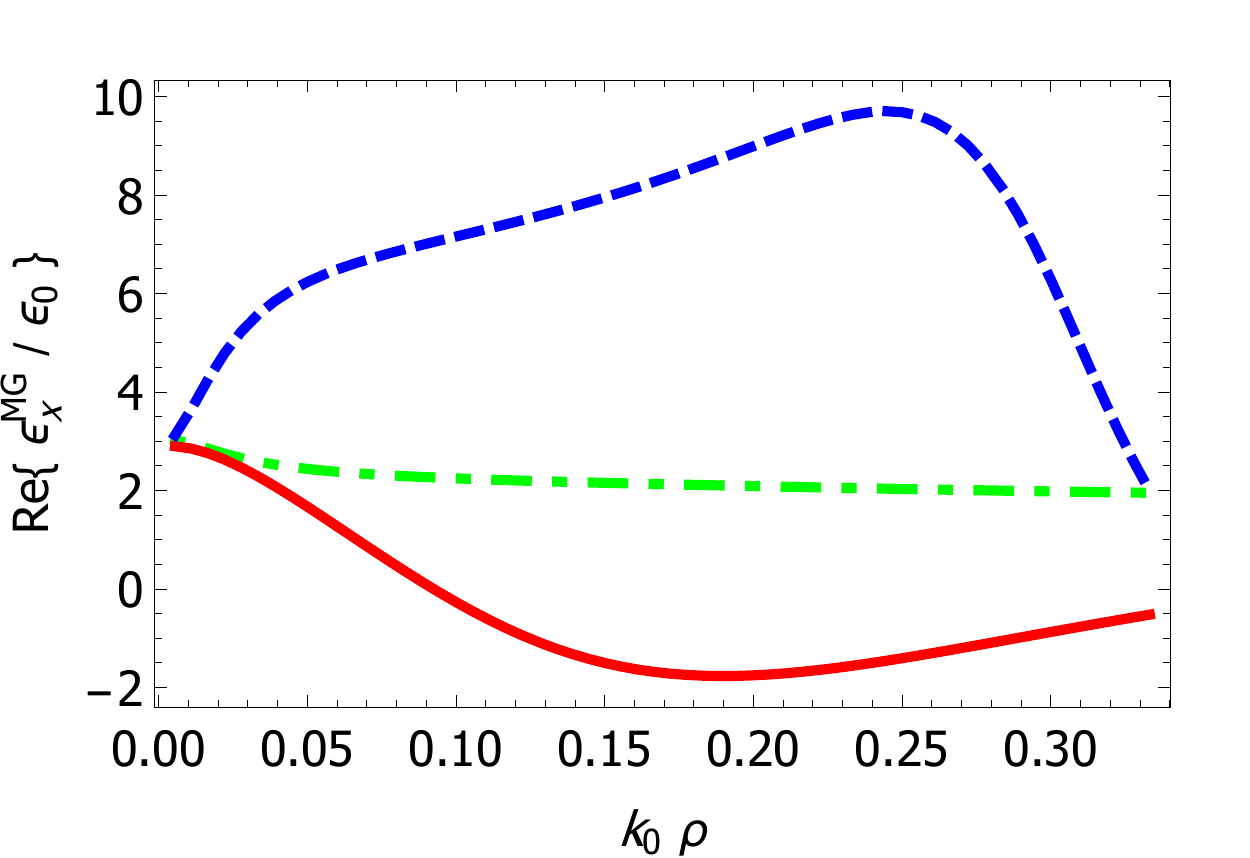} \hfill  \includegraphics[width=6.9cm]{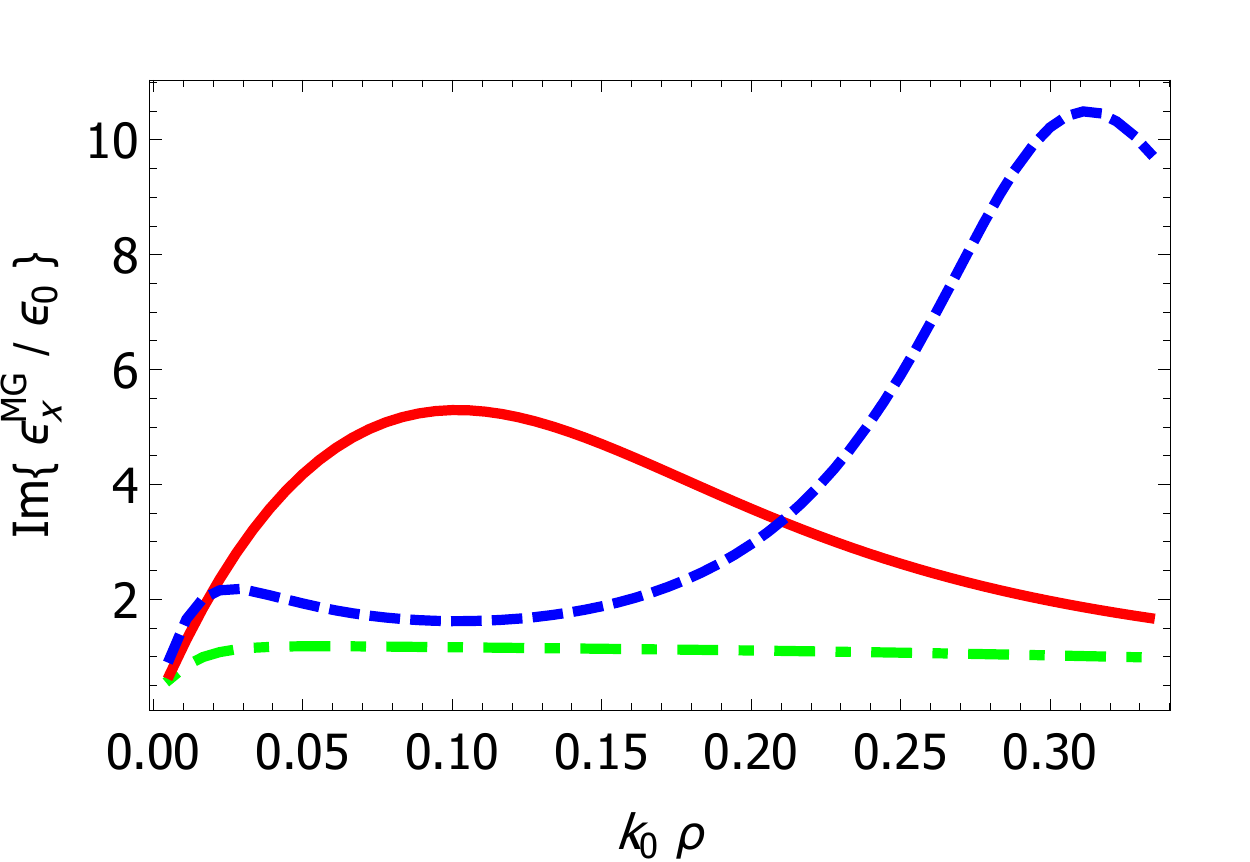}\\
  \includegraphics[width=6.9cm]{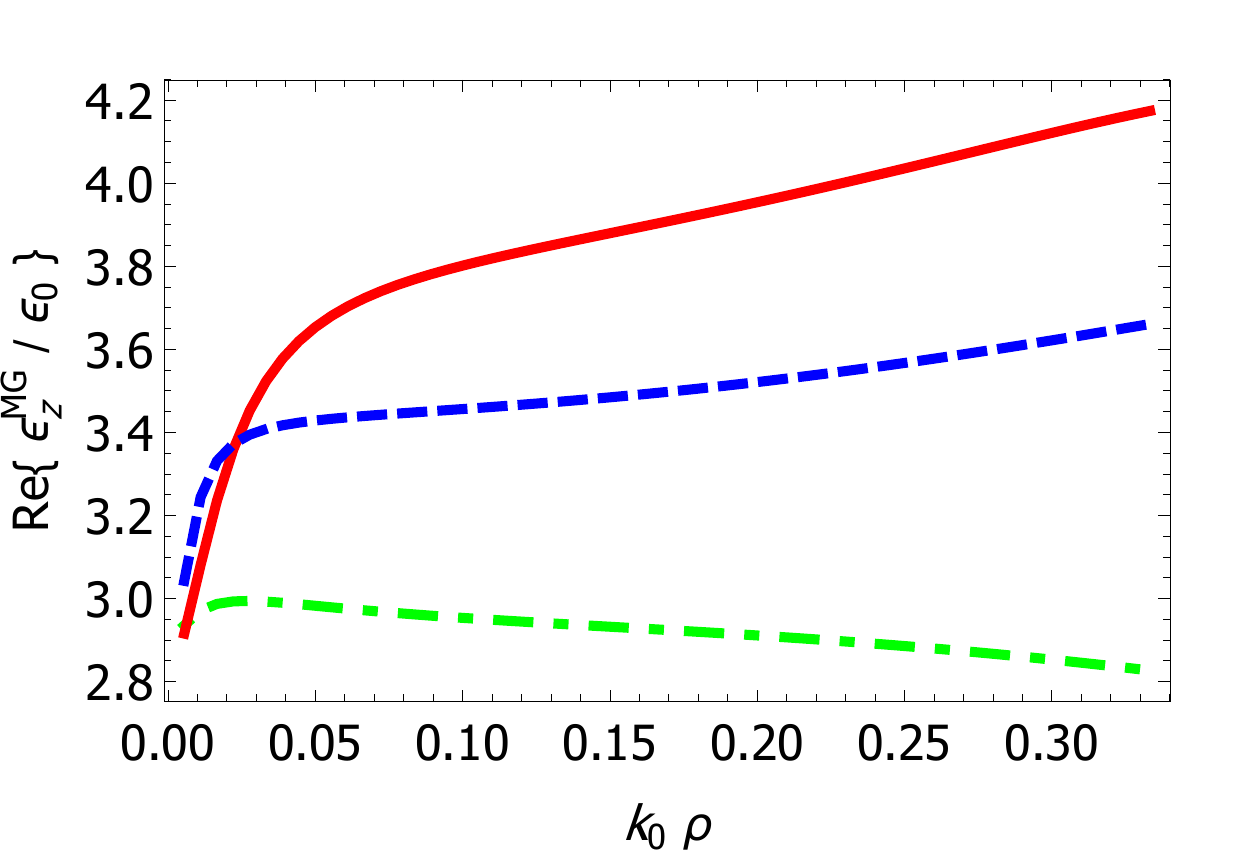} \hfill  \includegraphics[width=6.9cm]{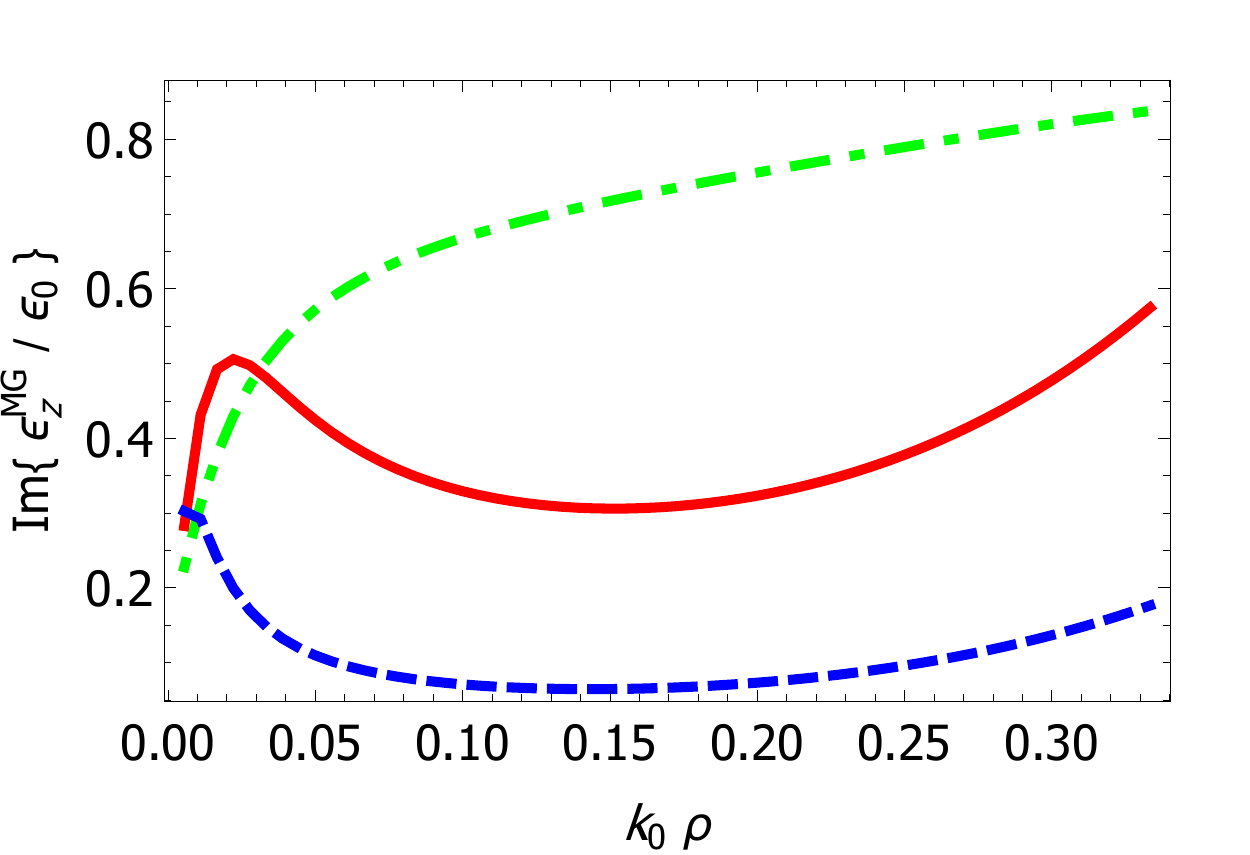}
 \caption{\label{Fig7}
 As Fig.~\ref{Fig2} except that $\underline{\underline{U}}$ is replaced by its orientational
average $\hat{\underline{\underline{U}}}$.}
\end{figure}

  Finally in this section, let us consider again the issue of orientation of the inclusions. The results presented in Figs.~\ref{Fig2}--\ref{Fig6} are based on the assumption that all inclusions  have the same orientation, as specified by the dyadic $\=S (\gamma, \beta, \psi)$. We now assume that all orientations of inclusions in the $xy$ plane are equally likely. Accordingly, the dyadic $\=U$  introduced in Eq.~\r{U_dyadic} is replaced by its orientational average 
  \begin{equation}
 \hat{\=U} = \frac{1}{2 \pi} \int^{2 \pi}_0 \=U \, d \psi,
  \end{equation}
  which yields
  \begin{equation}
 \hat{\=U} =  \rho \sqrt[3]{\le\frac{2}{U+1}\ri^2}\les \frac{U+1}{2} \le \ux \, \ux + \uy\,\uy \ri + \uz \, \uz \ris.
  \end{equation}
Since the symmetry axis of $\hat{\=U}$ is parallel to the optic axis of the vaterite host material, the permittivity dyadic of the  corresponding HCM has the uniaxial form
given in Eq.~\r{HCM_uniaxial}. 
Now we repeat the computations of Fig.~\ref{Fig2} but with $\=U$ replaced by its 
orientational
average $\hat{\=U}$. The corresponding real and imaginary parts of 
  the non-zero components of the HCM's relative permittivity dyadic   
 are plotted against relative size of gold nanoparticles
 in 
 Fig.~\ref{Fig7}. The plots of both the  real and imaginary parts of $ \eps^{MG}_z/\epso$ in Fig.~\ref{Fig7} are very similar to  those in Fig.~\ref{Fig2} for all wavelengths considered. But the plots
  of the  real and imaginary parts of $ \eps^{MG}_x/\epso$ in Fig.~\ref{Fig7} are both quite different to  those in Fig.~\ref{Fig2}; for example, the $\mbox{Re}\lec \eps^{MG}_x/\epso \ric$ plot in Fig.~\ref{Fig7} is concave whereas the corresponding
   plot  
 in Fig.~\ref{Fig2} is convex.
  
Parenthetically, homogenization formalisms such as the Maxwell Garnett and Bruggeman formalisms, as well as their extended variants, can accommodate other
 orientational statistics for $\=U$ in
 Eq.~\r{U_dyadic} than those considered here. Furthermore,  $\=U$ can be different for inclusions made of different materials
in these formalisms \c{MAEH}.

 \section{Closing remarks}
 
The non-zero components of the    permittivity dyadic of a
 biaxial dielectric
 HCM~---~constituted by porous vaterite impregnated with gold 
 nanoparticles~---~have been estimated using the extended Maxwell Garnett formalism. The extended formalism takes into account the anisotropy of the vaterite
 as well as the size, shape, and orientation of the nanoparticles and pores; and a size-dependent permittivity was used for the gold nanoparticles.
 Numerical studies revealed that the HCM's permittivity parameters are acutely sensitive to the size, shape, orientation, and volume fraction of the gold nanoparticles.
 Therefore, all of these attributes must be taken into account in theoretical studies of such HCMs.  In particular, when the size of the gold nanoparticles is neglected, the Maxwell Garnett formalism can provide quite different estimates of the HCM's permittivity parameters.
 
 Lastly, surface roughness of nanoparticles may also have a significant bearing on the permittivity dyadic of impregnated vaterite \c{Lu}. However, as 
 conventional approaches to homogenization such as 
 the Maxwell Garnett formalism are based on convex-shaped inclusions, surface roughness cannot be accommodated by such conventional approaches. This remains a matter for future research.
 
\section*{Appendix}

 In order to specify the size-dependent depolarization dyadic $\=D_{\,U,c}$ given in the extended Maxwell Garnett formula \r{epsMG}, it is convenient to adopt the
 most general linear framework in which case the host material $c$ is a bianisotropic material, characterized by four 3$\times$3 constitutive dyadics in the Tellegen formalism: the permittivity dyadic $\=\eps_{\,c}$, the permeability dyadic  $\=\mu_{\,c}$, and two magnetoelectric dyadics  $\=\xi_{\,c}$ and  $\=\zeta_{\,c}$. In the case of vaterite,  $\=\mu_{\,c} = \muo \=I$ and  $\=\xi_{\,c}  = \=\zeta_{\,c} = \=0$. 
 
 The size-dependent  6$\times$6 depolarization dyadic of an inclusion   embedded in a bianisotropic host material $c$ may be represented as the sum
 \begin{equation}
\*D_{\,U/\mbox{\tiny{c}}}   =   \*D^{
0}_{\,U/\mbox{\tiny{c}}}   +
\*D^{+}_{\,U/\mbox{\tiny{c}}}. \l{D_eta_def}
\end{equation}
 Herein the size-independent contribution \c{Ch3_M97,Ch3_MW97}
 \begin{equation}
\*D^{0}_{\,U/\mbox{\tiny{c}}}   = \frac{1}{4 \pi}  \int^{2 \pi}_{\phi = 0}
\int^{\pi}_{\theta = 0} \,  \underline{\underline{\check{\bf
G}}}^{\infty}_{\,\mbox{\tiny{c}}} (\=U^{-1}\.\hat{\#q}) 
 \sin \theta\, 
d\theta \, d \phi, \l{Ch3_depol_G_infty}
\end{equation}
 with the 6$\times$6 dyadic
 \begin{eqnarray}
&& \hspace{-20pt} \underline{\underline{\check{\bf
G}}}^{\infty}_{\,\mbox{\tiny{c}}} (\=U^{-1}\.\hat{\#q}) =
\displaystyle{\frac{1 }{i \omega \, b_{\,U/\mbox{\tiny{c}}}(\theta, \phi)}} \times
 \nonumber \\ && \hspace{-20pt} 
\les
\begin{array}{cc}
\beta_\mu (\theta, \phi) \; \=U^{-1}\. \hat{\#q} \hat{\#q} \. \=U^{-1}  &
-\beta_\xi (\theta, \phi) \; \=U^{-1}\. \hat{\#q} \hat{\#q}\. \=U^{-1} \vspace{4pt}\\
-\beta_\zeta (\theta, \phi) \; \=U^{-1}\.  \hat{\#q} \hat{\#q} \. \=U^{-1}&
\beta_\eps (\theta, \phi) \; \=U^{-1}\. \hat{\#q} \hat{\#q} \. \=U^{-1}
\end{array}
\ris
 ;  \nonumber \\ &&
 \end{eqnarray}
 the
 scalar functions
 \begin{equation}
\left.
\begin{array}{l}
\beta_\sigma (\theta, \phi) = 
 \hat{\#q}\. \=U^{-1} \. \={\sigma}_{\,\mbox{\tiny{c}}}
\. \=U^{-1}\.\hat{\#q}, \quad \le \sigma \in \lec \eps, \xi, \zeta, \mu\ric \ri \vspace{8pt}\\
b_{\,U/\mbox{\tiny{c}}}(\theta, \phi) = \les \beta_\eps (\theta, \phi)
\, \beta_\mu (\theta, \phi) \ris - \les \beta_\xi (\theta, \phi) \,
\beta_\zeta (\theta, \phi) \ris
\end{array}
\right\};
\end{equation}
and the unit vector 
\begin{equation}
\hat{\#q} = \frac{1}{q} \,\#q = \sin \theta \cos \phi \, \ux \,
+ \sin \theta \sin \phi \,\uy \,  + \cos \theta \, \uz \,  .
\end{equation}
 The size-dependent contribution is \c{Ch3_M_WRM,Ch3_JJ_WRCM}
 \begin{eqnarray} \l{D_plus}
\*D^{+}_{\,U/\mbox{\tiny{c}}}   &=&
\frac{\omega^4}{4 \pi } \int^{2 \pi}_{\phi = 0} \int^{\pi}_{\theta =
0} \frac{1}{b_{\,U/\mbox{\tiny{c}}} ( \theta, \phi )
}\Bigg[ \frac{1}{ \kappa_+  - \kappa_-  }  \Bigg( \frac{\exp \le i \rho q \ri }{2 q^2}
\nonumber \\ &&
\times \le 1 - i \rho q\ri
 \Big\{ \,
 \mbox{det} \les \underline{\underline{\check{\bf
A}}}_{\,\mbox{\tiny{c}}}(\=U^{-1}\.\#q) \ris  \, \underline{\underline{\check{\bf
G}}}^{+}_{\,\mbox{\tiny{c}}}(\=U^{-1}\.\#q)\nonumber \\
&&
  +  \mbox{det} \les \underline{\underline{\check{\bf
A}}}_{\,\mbox{\tiny{c}}}(-\=U^{-1}\.\#q) \ris \,  \underline{\underline{\check{\bf
G}}}^{+}_{\,\mbox{\tiny{c}}}(-\=U^{-1}\.\#q) \Big\}
\Bigg)^{q= \sqrt{ \kappa_+}}_{q= \sqrt{\kappa_-}}  
\nonumber \\ &&
+ \frac{
 \mbox{det} \les \underline{\underline{\check{\bf
A}}}_{\,\mbox{\tiny{c}}}(\#0) \ris} {\kappa_+  \,
\kappa_- }\, \underline{\underline{\check{\bf
G}}}^{+}_{\,\mbox{\tiny{c}}}(\#0)
 \Bigg]\, \sin
\theta \; d \theta \; d \phi,  \l{D_e}
\end{eqnarray}
with the 6$\times$6 dyadics
\begin{equation}
\left.
\begin{array}{l}
\displaystyle{
\underline{\underline{\check{\bf A}}}_{\,\mbox{\tiny{c}}}(\#q) = 
 \les
\begin{array}{cc} \=0 &  (\#q /\omega) \times \=I \\ \vspace{-6pt}
\\
  -(\#q /\omega) \times \=I & \=0
\end{array} \ris +
\les 
\begin{array}{cc} \=\eps_{\,c} & \=\xi_{\,c}\\ \=\zeta_{\,c} & \=\mu_{\,c} \end{array}
\ris} \vspace{8pt} \\
\displaystyle{\underline{\underline{\check{\bf
G}}}^{+}_{\,\mbox{\tiny{c}}}(\#q) =\frac{1}{i \omega}\, \underline{\underline{\check{\bf A}}}^{-1}_{\,\mbox{\tiny{c}}}(\#q)
- \underline{\underline{\check{\bf
G}}}^{\infty}_{\,\mbox{\tiny{c}}} (\=U^{-1}\.\hat{\#q})}
\end{array} \right\}
.
\end{equation}
and
$\kappa_\pm $ representing the  roots of the quadratic (in $q^2$) equation $\mbox{det} \les
\underline{\underline{\check{\bf
A}}}_{\,\mbox{\tiny{c}}}(\=U^{-1}\.\#q) \ris = 0$, i.e.,
\begin{equation}
\mbox{det} \les
\underline{\underline{\check{\bf
A}}}_{\,\mbox{\tiny{c}}}(\=U^{-1}\.\#q) \ris \propto \le q^2 - \kappa_+ \ri \le q^2 - \kappa_- \ri = 0.
\end{equation}
Thus, the desired 3$\times$3 depolarization dyadic  $\=D_{\,U,c}$
 arises as
 \begin{equation}
 \l{Dee_etc}
 \=D_{\,U,c}
= \les \, \begin{array}{ccc}
\les \*D_{\,U/\mbox{\tiny{c}}}  \ris_{1,1} & 
\les \*D_{\,U/\mbox{\tiny{c}}}  \ris_{1,2} & 
\les \*D_{\,U/\mbox{\tiny{c}}}  \ris_{1,3} \vspace{4pt} \\
\les \*D_{\,U/\mbox{\tiny{c}}}  \ris_{2,1} & 
\les \*D_{\,U/\mbox{\tiny{c}}}  \ris_{2,2} & 
\les \*D_{\,U/\mbox{\tiny{c}}}  \ris_{2,3} \vspace{4pt} \\
\les \*D_{\,U/\mbox{\tiny{c}}}  \ris_{3,1} & 
\les \*D_{\,U/\mbox{\tiny{c}}}  \ris_{3,2} & 
\les \*D_{\,U/\mbox{\tiny{c}}}  \ris_{3,3} 
\end{array} \, \ris.
\end{equation}

In general, numerical methods are needed to evaluate the double integrals in Eqs.~\r{Ch3_depol_G_infty} and \r{D_plus}, but for biaxial dielectric host materials those
integrals in Eqs.~\r{Ch3_depol_G_infty} may be expressed in terms of incomplete elliptic integrals of the first and second kind \c{Ch3_W98b}.

 \vspace{5mm}

\noindent {\bf Acknowledgments.}
This work was supported  in part by
 EPSRC (grant number EP/V046322/1) and  US NSF (grant number DMS-2011996).
AL thanks the Charles Godfrey Binder Endowment at The Pennsylvania State University    for ongoing support of his research endeavors.

\end{document}